\documentclass[twocolumn,showpacs,preprintnumbers,amsmath,amssymb,superscriptaddress, prd]{revtex4-1}

\usepackage{graphicx}
\usepackage{dcolumn}
\usepackage{bm}
\usepackage{color}
\usepackage{lineno}
\usepackage{amsmath}
\usepackage{relsize}
\usepackage{tabularx}
\usepackage{relsize}

\begin{document}

\title{Differential limit on the extremely-high-energy cosmic neutrino flux in the presence
  of astrophysical background from nine years of IceCube data}
  
\affiliation{III. Physikalisches Institut, RWTH Aachen University, D-52056 Aachen, Germany}
\affiliation{Department of Physics, University of Adelaide, Adelaide, 5005, Australia}
\affiliation{Dept.~of Physics and Astronomy, University of Alaska Anchorage, 3211 Providence Dr., Anchorage, AK 99508, USA}
\affiliation{Dept.~of Physics, University of Texas at Arlington, 502 Yates St., Science Hall Rm 108, Box 19059, Arlington, TX 76019, USA}
\affiliation{CTSPS, Clark-Atlanta University, Atlanta, GA 30314, USA}
\affiliation{School of Physics and Center for Relativistic Astrophysics, Georgia Institute of Technology, Atlanta, GA 30332, USA}
\affiliation{Dept.~of Physics, Southern University, Baton Rouge, LA 70813, USA}
\affiliation{Dept.~of Physics, University of California, Berkeley, CA 94720, USA}
\affiliation{Lawrence Berkeley National Laboratory, Berkeley, CA 94720, USA}
\affiliation{Institut f\"ur Physik, Humboldt-Universit\"at zu Berlin, D-12489 Berlin, Germany}
\affiliation{Fakult\"at f\"ur Physik \& Astronomie, Ruhr-Universit\"at Bochum, D-44780 Bochum, Germany}
\affiliation{Universit\'e Libre de Bruxelles, Science Faculty CP230, B-1050 Brussels, Belgium}
\affiliation{Vrije Universiteit Brussel (VUB), Dienst ELEM, B-1050 Brussels, Belgium}
\affiliation{Dept.~of Physics, Massachusetts Institute of Technology, Cambridge, MA 02139, USA}
\affiliation{Dept. of Physics and Institute for Global Prominent Research, Chiba University, Chiba 263-8522, Japan}
\affiliation{Dept.~of Physics and Astronomy, University of Canterbury, Private Bag 4800, Christchurch, New Zealand}
\affiliation{Dept.~of Physics, University of Maryland, College Park, MD 20742, USA}
\affiliation{Dept.~of Physics and Center for Cosmology and Astro-Particle Physics, Ohio State University, Columbus, OH 43210, USA}
\affiliation{Dept.~of Astronomy, Ohio State University, Columbus, OH 43210, USA}
\affiliation{Niels Bohr Institute, University of Copenhagen, DK-2100 Copenhagen, Denmark}
\affiliation{Dept.~of Physics, TU Dortmund University, D-44221 Dortmund, Germany}
\affiliation{Dept.~of Physics and Astronomy, Michigan State University, East Lansing, MI 48824, USA}
\affiliation{Dept.~of Physics, University of Alberta, Edmonton, Alberta, Canada T6G 2E1}
\affiliation{Erlangen Centre for Astroparticle Physics, Friedrich-Alexander-Universit\"at Erlangen-N\"urnberg, D-91058 Erlangen, Germany}
\affiliation{D\'epartement de physique nucl\'eaire et corpusculaire, Universit\'e de Gen\`eve, CH-1211 Gen\`eve, Switzerland}
\affiliation{Dept.~of Physics and Astronomy, University of Gent, B-9000 Gent, Belgium}
\affiliation{Dept.~of Physics and Astronomy, University of California, Irvine, CA 92697, USA}
\affiliation{Dept.~of Physics and Astronomy, University of Kansas, Lawrence, KS 66045, USA}
\affiliation{SNOLAB, 1039 Regional Road 24, Creighton Mine 9, Lively, ON, Canada P3Y 1N2}
\affiliation{Department of Physics and Astronomy, UCLA, Los Angeles, CA 90095, USA}
\affiliation{Dept.~of Astronomy, University of Wisconsin, Madison, WI 53706, USA}
\affiliation{Dept.~of Physics and Wisconsin IceCube Particle Astrophysics Center, University of Wisconsin, Madison, WI 53706, USA}
\affiliation{Institute of Physics, University of Mainz, Staudinger Weg 7, D-55099 Mainz, Germany}
\affiliation{Department of Physics, Marquette University, Milwaukee, WI, 53201, USA}
\affiliation{Physik-department, Technische Universit\"at M\"unchen, D-85748 Garching, Germany}
\affiliation{Institut f\"ur Kernphysik, Westf\"alische Wilhelms-Universit\"at M\"unster, D-48149 M\"unster, Germany}
\affiliation{Bartol Research Institute and Dept.~of Physics and Astronomy, University of Delaware, Newark, DE 19716, USA}
\affiliation{Dept.~of Physics, Yale University, New Haven, CT 06520, USA}
\affiliation{Dept.~of Physics, University of Oxford, 1 Keble Road, Oxford OX1 3NP, UK}
\affiliation{Dept.~of Physics, Drexel University, 3141 Chestnut Street, Philadelphia, PA 19104, USA}
\affiliation{Physics Department, South Dakota School of Mines and Technology, Rapid City, SD 57701, USA}
\affiliation{Dept.~of Physics, University of Wisconsin, River Falls, WI 54022, USA}
\affiliation{Dept.~of Physics and Astronomy, University of Rochester, Rochester, NY 14627, USA}
\affiliation{Oskar Klein Centre and Dept.~of Physics, Stockholm University, SE-10691 Stockholm, Sweden}
\affiliation{Dept.~of Physics and Astronomy, Stony Brook University, Stony Brook, NY 11794-3800, USA}
\affiliation{Dept.~of Physics, Sungkyunkwan University, Suwon 440-746, Korea}
\affiliation{Dept.~of Physics and Astronomy, University of Alabama, Tuscaloosa, AL 35487, USA}
\affiliation{Dept.~of Astronomy and Astrophysics, Pennsylvania State University, University Park, PA 16802, USA}
\affiliation{Dept.~of Physics, Pennsylvania State University, University Park, PA 16802, USA}
\affiliation{Dept.~of Physics and Astronomy, Uppsala University, Box 516, S-75120 Uppsala, Sweden}
\affiliation{Dept.~of Physics, University of Wuppertal, D-42119 Wuppertal, Germany}
\affiliation{DESY, D-15738 Zeuthen, Germany}

\author{M.~G.~Aartsen}
\affiliation{Dept.~of Physics and Astronomy, University of Canterbury, Private Bag 4800, Christchurch, New Zealand}
\author{M.~Ackermann}
\affiliation{DESY, D-15738 Zeuthen, Germany}
\author{J.~Adams}
\affiliation{Dept.~of Physics and Astronomy, University of Canterbury, Private Bag 4800, Christchurch, New Zealand}
\author{J.~A.~Aguilar}
\affiliation{Universit\'e Libre de Bruxelles, Science Faculty CP230, B-1050 Brussels, Belgium}
\author{M.~Ahlers}
\affiliation{Niels Bohr Institute, University of Copenhagen, DK-2100 Copenhagen, Denmark}
\author{M.~Ahrens}
\affiliation{Oskar Klein Centre and Dept.~of Physics, Stockholm University, SE-10691 Stockholm, Sweden}
\author{I.~Al~Samarai}
\affiliation{D\'epartement de physique nucl\'eaire et corpusculaire, Universit\'e de Gen\`eve, CH-1211 Gen\`eve, Switzerland}
\author{D.~Altmann}
\affiliation{Erlangen Centre for Astroparticle Physics, Friedrich-Alexander-Universit\"at Erlangen-N\"urnberg, D-91058 Erlangen, Germany}
\author{K.~Andeen}
\affiliation{Department of Physics, Marquette University, Milwaukee, WI, 53201, USA}
\author{T.~Anderson}
\affiliation{Dept.~of Physics, Pennsylvania State University, University Park, PA 16802, USA}
% ...

\date{\today}

\collaboration{IceCube Collaboration}
\thanks{Email: analysis@icecube.wisc.edu}
\noaffiliation



\date{\today}

\begin{abstract}
We report a quasi-differential upper limit on the extremely-high-energy (EHE)
neutrino flux above $5\times 10^{6}$ GeV based on an analysis of nine years of IceCube data.
The astrophysical neutrino flux measured by IceCube extends to PeV energies,
and it is a background flux when searching for an independent signal flux
at higher energies, such as the cosmogenic neutrino signal.
We have developed a new method to place robust limits
on the EHE neutrino flux in the presence of an astrophysical background,
whose spectrum has yet to be understood with high precision at PeV energies.
A distinct event with a deposited energy above $10^{6}$ GeV was found in the new two-year sample,
in addition to the one event previously found in the seven-year EHE neutrino search.
These two events represent a neutrino flux that is incompatible with predictions for a cosmogenic neutrino flux
and are considered to be an astrophysical background in the current study.
The obtained limit is the most stringent to date in the energy range
between $5 \times 10^{6}$ and $2 \times 10^{10}$ GeV.
This result constrains neutrino models
predicting a three-flavor neutrino flux of
$E_\nu^2\phi_{\nu_e+\nu_\mu+\nu_\tau}\simeq2\times 10^{-8}\ {\rm GeV}/{\rm cm}^2\ \sec\ {\rm sr}$
at $10^9\ {\rm GeV}$.
A significant part of the parameter space for EHE neutrino production scenarios
assuming a proton-dominated composition of ultra-high-energy cosmic rays is 
disfavored independently of uncertain models of the extragalactic background light
which previous IceCube constraints partially relied on.
\end{abstract}

\maketitle

\section{\label{sec:intro} Introduction}
The origin of ultra-high-energy cosmic rays (UHECRs; cosmic rays with energies
greater than about $10^{18}$~eV) is among the long-standing questions in astrophysics.
Recent measurements indicate that they originate from extragalactic sources~\cite{augerScience2017}.
Secondary extremely-high-energy (EHE) neutrinos produced by UHECR interactions with background radiation
provide an alternative and promising indicator of UHECR sources
as neutrinos propagate cosmological distances without interaction or deflection by magnetic fields.
A series of EHE neutrino searches have been conducted~\cite{EHE2011, EHE2012,augerNu2015,EHE2016},
however, cosmogenic neutrinos induced by the Greisen--Zatsepin--Kuzmin
(GZK) mechanism~\cite{berezinsky69} have not been detected.
Because the cosmogenic neutrino rates strongly depend on the UHECR source evolution function
that characterizes the source classes~\cite{yoshidaIshihara2012,Kotera2010, Decerprit2011},
recent limits on the EHE neutrino flux by the IceCube Neutrino Observatory have provided
a unique constraint on UHECR sources. The aforementioned limits published by IceCube~\cite{EHE2016}
and subsequently reported by Auger~\cite{augerICRC2017} indicate that objects 
with a cosmological evolution stronger than the star formation rate (SFR) are
disfavored as UHECR sources, if the UHECRs are proton-dominated.

A differential limit is an effective way to characterize the energy dependence of an experiment's sensitivity.
As each experiment is sensitive to neutrinos of different energy,
a model-dependent constraint does not indicate which energy region contributes most to
bounding a given model.
In the case of the null observation, Anchordoqui {\it et al.}~\cite{Anchordoqui} proposed
setting a quasi-differential limit:
\begin{equation}
  \phi_{\nu_e+\nu_\mu+\nu_\tau}^{\rm UL}(E_\nu)=
  3\frac{\rm N_{90}}{4\pi E_\nu T{\rm log 10} \sum_{i=\nu_e,\nu_\mu,\nu_\tau}A^\nu_i(E_\nu)},
      \label{eq:dif_limit_analytical}
\end{equation}
where $T$ is the observation time, $A^\nu_i$ is the $4\pi$-averaged neutrino
effective area for a neutrino flavor $i$, and ${\rm N_{90}}$ is the 90\% C.L. upper limit on the number of events.
The Feldman--Cousins method~\cite{feldman98} sets ${\rm N_{90}} = 2.4$ for the case of negligible background.
An equal flavor ratio of neutrino fluxes $\nu_e : \nu_\mu : \nu_\tau = 1:1:1$ at the Earth is assumed.
This upper limit of Eq.~(\ref{eq:dif_limit_analytical})
is equivalent to the limit on the normalization
of neutrino fluxes following $E_\nu^{-1}$ with an interval of one decade.

This formula must be modified
when neutrino event candidates are contained in the data sample.
However, it is not clear what approach should be employed to incorporate detected events
in the calculation of the differential limit.
Muon neutrino events deposit an unmeasured fraction of their energy
outside the instrument volume.  Therefore, a large uncertainty in the measured
muon neutrino energy cannot be avoided.
The probability density function (PDF) of the observed neutrino energy
thus depends on the as-yet unknown true neutrino energy spectrum.
In the EHE neutrino analysis with IceCube published in 2013~\cite{EHE2012}, in which the first PeV events
were detected in two years of data~\cite{icecubePeV2013},
the upper limit on the number of events $N_{90}$ in Eq.~(\ref{eq:dif_limit_analytical})
was derived from the probability of finding $n$ ($n=0,1,2,...$) neutrino events
in an interval of one decade:
$[\log_{10}(E_\nu/{\rm GeV})-0.5,\log_{10}(E_\nu/{\rm GeV})+0.5]$.
This probability was estimated using the PDF of the primary neutrino energy
for each of the detected events, assuming the parent neutrino energy spectrum followed $E_\nu^{-2}$.
However, the confidence coverage is not well defined in this approach
as each of the Poisson upper limits at 90 \% confidence level in case of finding $n$ events are further weighted
by the $n$ event detection Bayesian probability. Moreover, it
does not fully consider the energy dependence of the background contamination.

In this paper, we present a complete frequentist approach to calculate the
flux limits and update the constraints
using a collection of IceCube data taken over nine years from
April 2008 to May 2017.
The data sample includes two additional years of IceCube data in addition to the seven-year sample used
in the previous EHE analysis~\cite{EHE2016}.
All signal selection criteria are the same as in the previous publication and
described in Secs. \ref{sec:data} and \ref{sec:selection}.
The new approach using a nuisance parameter
to represent the unknown astrophysical background
and the method of p-value calculations using the Poisson-binned likelihood ratio are
presented in Sec. \ref{sec:likelihood}.
Last, the results and implications
of the derived limits for explaining the origin of UHECRs are discussed in Sec. \ref{sec:results}.

\section{\label{sec:data} Data and Simulation}

IceCube is a cubic-kilometer neutrino detector installed in the ice under the South Pole
between depths of 1450 m and 2450 m, forming a three-dimensional array of
digital optical modules (DOMs)~\cite{IceCubeDetector}.
To form the detector, cable assemblies called strings were lowered into holes drilled vertically
into the glacier ice with a horizontal spacing of approximately 125~m.
The detector construction was completed
in December 2010 and the observatory has been in full operation with 86 strings (IC86) since
May 2011. During the construction period, it was partially operated with 40, 59, and 79 strings
in 2008--2009, 2009--2010, and 2010--2011, respectively. The analysis described here
is based on data taken from April 2008 to May 2017. The effective live time of the sample
is 3142.5 days. The most recent two-year data sample provides approximately 30\% more
exposure than the previous EHE neutrino search~\cite{EHE2016}.

There are two classes of atmospheric background events:
atmospheric muon bundles and
events generated by atmospheric neutrinos.
They were simulated using the CORSIKA~\cite{corsika} package with the SIBYLL
hadronic interaction model~\cite{Ahn:2009wx}
and by the IceCube neutrino-generator program based on the ANIS code~\cite{anis}, respectively.
Prompt atmospheric neutrinos from short-lived heavy meson decays
were modeled following~\cite{Enberg2008}, which predicted a higher prompt neutrino flux than recent
calculations~\cite{Bhattacharya2016}, and represent a conservative background estimate.
The EHE neutrino-induced events were simulated by the JULIeT package~\cite{juliet},
which provides the cosmogenic (GZK) signal simulation sample as well as
simulations of the astrophysical {\it background} events, whose spectrum is assumed to
be described by an unbroken power law in the relevant energy region.
The detailed simulation procedure used in this work is described in Ref.~\cite{EHE2012}.

\section{\label{sec:selection} Event Selection}

\begin{figure*}
  \includegraphics[width=0.4\textwidth]{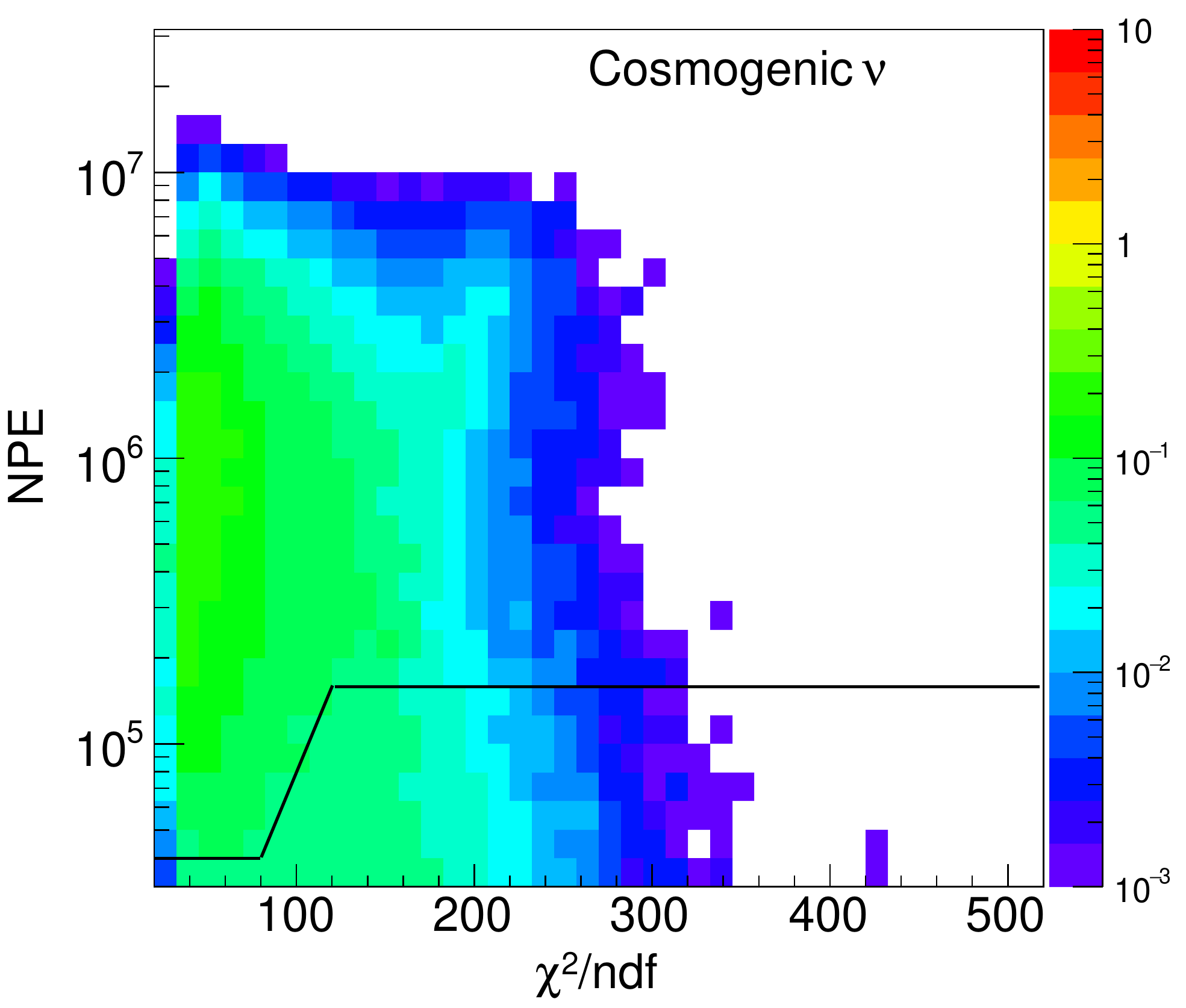}
  \includegraphics[width=0.4\textwidth]{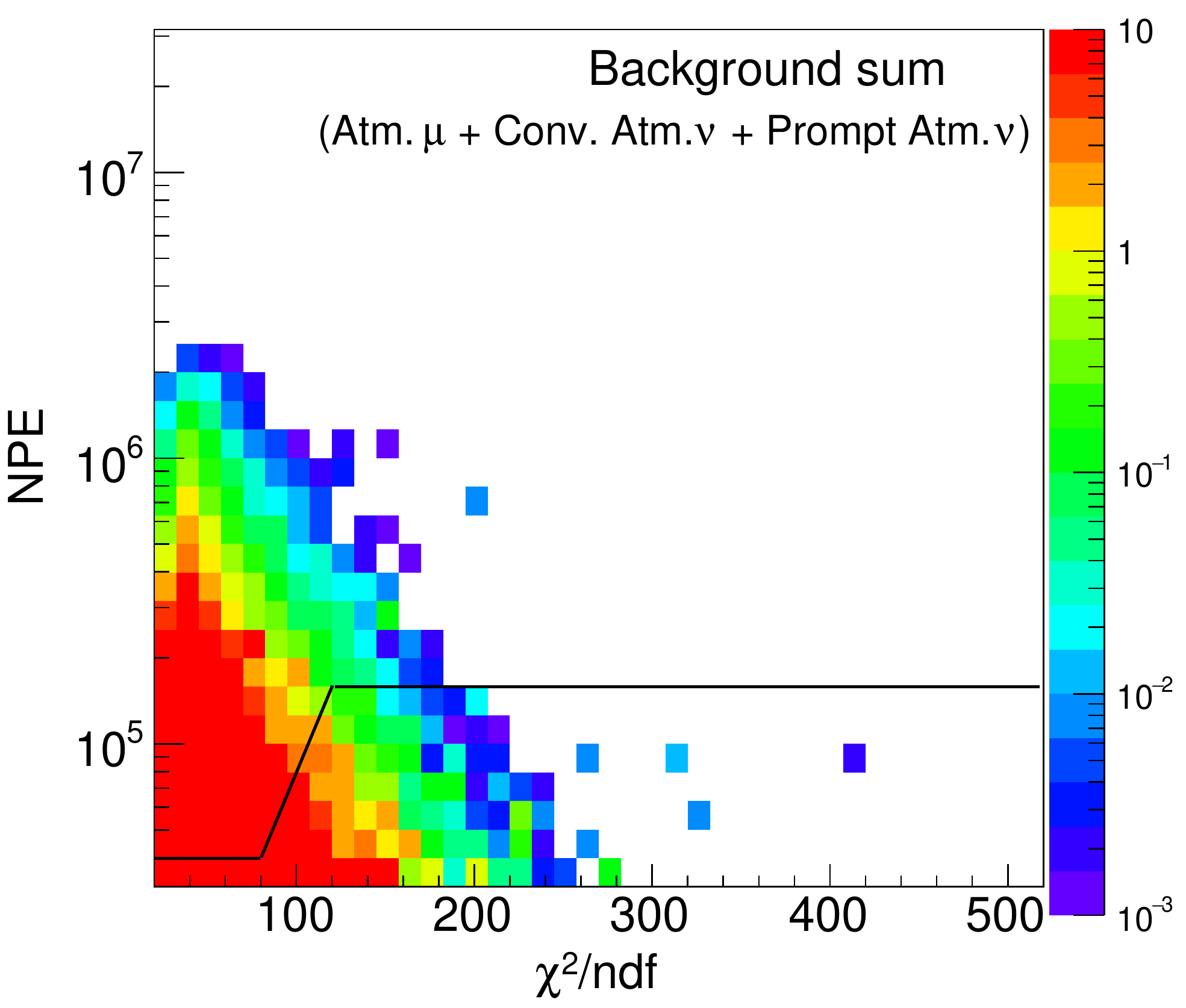}
  \caption{Event count distributions before the track quality cut of the sample,
    including all three flavors of neutrinos as a function of NPE and $\chi^2_{\rm track}/ndf$.
    The colors indicate the expected number of events seen by the IceCube
    EHE neutrino analysis in the nine-year exposure.
    The solid line in each panel indicates the track quality selection criteria,
    where events above the lines are retained. Simulations of a cosmogenic (GZK)
    neutrino model~\cite{Ahlers2010} are shown in the left panel, and the background simulations
    of atmospheric muons, conventional atmospheric neutrinos, and prompt atmospheric neutrinos
    are shown in the right panel.}
  \label{fig:TrackQualityCut}
\end{figure*}

\begin{figure*}
  \includegraphics[width=0.4\textwidth]{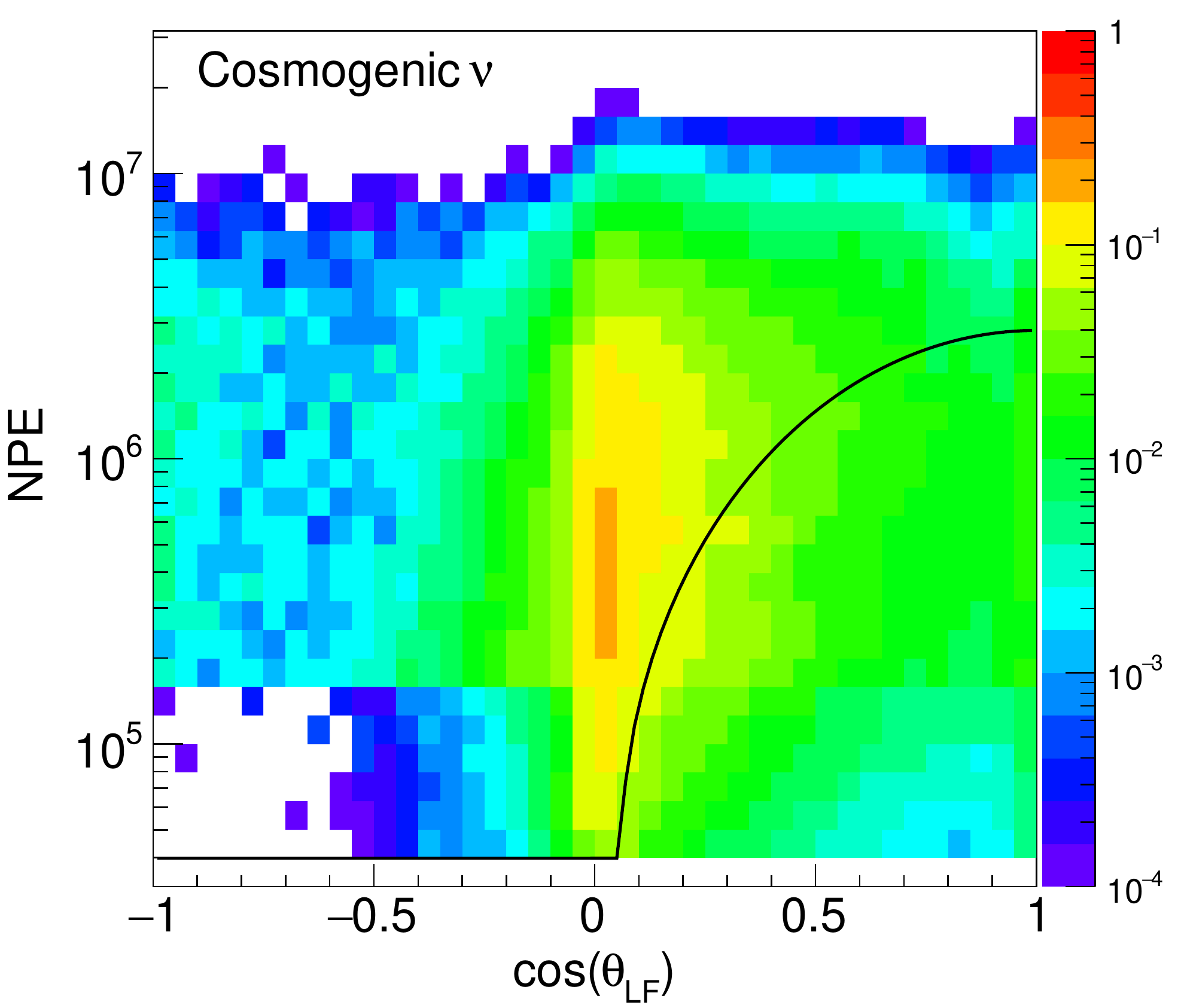}
  \includegraphics[width=0.4\textwidth]{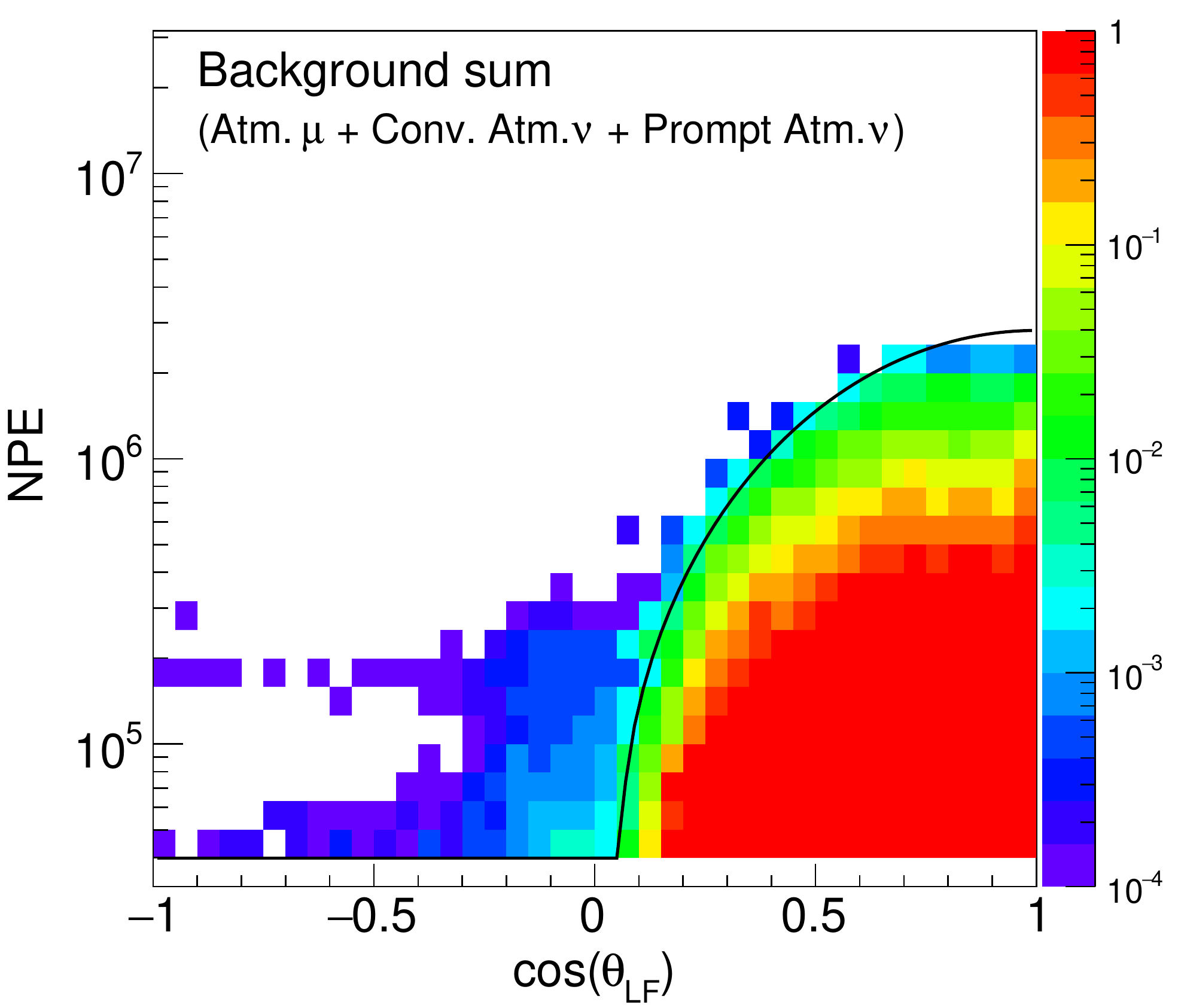}
  \caption{Event count distributions before the muon bundle cut,
    including all three flavors of neutrinos as a function of NPE and $\cos(\theta_{\rm LF})$.
    The colors indicate the expected number of events seen by the IceCube
    EHE neutrino analysis in the nine-year exposure.
    The solid line in each panel indicates the muon bundle selection criteria,
    where events above the lines are retained.  The major sources of remaining background
    are rare high-energy atmospheric neutrinos and muons originating in UHECRs.  Again, cosmogenic (GZK)
    neutrinos are shown in the left panel, while background simulations are shown in the right panel. }
  \label{fig:MuonBundleCut}
\end{figure*}
The EHE signal selection criteria remain the same as in the previous analysis~\cite{EHE2016}.
The selection criteria were determined by following a blind analysis strategy,
and the cut value optimization was carried out by looking at the simulated event samples with
the experimental data blind, except for a 10\% subset
of experimental data used to validate the simulation.
The backgrounds for the EHE neutrino search
are atmospheric muon bundles and atmospheric neutrinos
initiated in cosmic-ray air showers.
As the EHE signal events deposit more energy in the form of Cherenkov light than the background,
the total number of photoelectrons (NPE) recorded in an event is used as
the main distinctive feature to eliminate the background. This basic algorithm
was established in the EHE neutrino search based on two years of IceCube data~\cite{EHE2012}.

\begin{table*}
  \caption{Rates and fractions of simulated data surviving by type
    as a function of event selection level applied with IC86 configuration.
    Efficiencies are calculated with respect to the online EHE filter.}
  \label{tbl:passing_rates}
\begin{center}
\begin{tabular}{c@{\hspace{1.0cm}}c@{\hspace{0.8cm}}c@{\hspace{0.8cm}}c}
\hline
\hline
Cut level & atmospheric muons       & atmospheric neutrinos    & signal cosmogenic neutrinos~\cite{Ahlers2010} \\
          & number [Hz]& number [Hz]  & fraction surviving (\%) \\
\hline
\hline
Online EHE Filter & $0.8$               & $7.6\times 10^{-6}$ & 100 \\
Offline EHE Cut   & $6.7\times 10^{-4}$   & $1.0\times 10^{-8}$ &  74 \\
Track Quality Cut & $1.6\times 10^{-4}$   & $6.1\times 10^{-10}$ &  61 \\
Muon bundle Cut   & $3.0\times 10^{-10}$  & $3.6\times 10^{-10}$ &  43 \\
\hline
\hline
\end{tabular}
\end{center}
\end{table*}

The {\it Online EHE Filter} first selects events with an NPE greater than 1000 photoelectrons (p.e).
After removing DOM signals from coincident atmospheric muons and
photomultiplier tube dark noise~\cite{EHE2012},
the {\it Offline EHE Cut} selects candidate events by requesting at least
25,000 p.e. and more than 100 hit DOMs.
The technical details such as the NPE extraction method
and hit cleaning algorithm are fully described in Ref.~\cite{EHE2012}.
The event direction of surviving events is reconstructed by the LineFit algorithm~\cite{lineFit}
that masks photon hits which have substantially different timing distributions
from Cherenkov photons radiated by an EHE muon track~\cite{EHElineFit}.

The {\it Track Quality Cut} is then applied based on
the LineFit goodness-of-fit parameter $\chi^2_{\rm track}/ndf$,
which is a measure of the consistency with a track-like event topology.
Track-like events (primarily from muons and EHE taus) generally yield smaller NPE
than cascade-like events (primarily from  electrons and hadrons) of the same energy
as track-like events deposit only a small fraction of the parent neutrino energy
within the detection volume.  Consequently,
we reduced the NPE threshold for track-like events and
relative to the cascade-like events in the track quality cut:
\begin{equation}
  {\small
  \log_{10}{{\rm NPE}}  \geq
  \left\{
  \begin{array}{lc}
    4.6 &  {(\small \frac{\chi^2_{\rm track}}{ndf}< 80)}\\
    0.015 \left(\frac{\chi^2_{\rm track}}{ndf}-80\right)+4.6 &  \\
        &  {(\small 80\leq \frac{\chi^2_{\rm track}}{ndf}< 120)}\\
    5.2 &  {(\small 120\leq \frac{\chi^2_{\rm track}}{ndf})}.
  \end{array}
  \right.
  }
  \label{eq:track_cut}
\end{equation}
Figure~\ref{fig:TrackQualityCut} shows the signal and background event distributions
as a function of $\chi^2_{\rm track}/ndf$ and NPE. The solid line represents
the cut described by Eq.~\ref{eq:track_cut}.
Note that this selection criterion filters out the previously observed PeV energy
neutrino-induced cascade events~\cite{icecubePeV2013}.
Muon track events dominate for $\chi^2_{\rm track}/ndf<80$. A subsample of events
that meets this $\chi^2_{\rm track}$ condition
is used for the EHE track alert system~\cite{icecubeRealtime}.

The final event selection cut is made based on the
NPE and LineFit reconstructed zenith angle ($\cos\theta_{\rm LF}$).
A zenith angle-dependent NPE threshold is used to remove the atmospheric muon background
in the downward-going region. The selection criteria in this {\it Muon Bundle Cut} are
\begin{equation}
  {\small
  \log_{10}{{\rm NPE}} \geq \left\{
  \begin{array}{ll}
    4.6 & {(\small \cos\theta_{\rm LF}< 0.06)}\\
    4.6+1.85\times & \\
    \sqrt{1.0-\left(\frac{1.0-\cos\theta_{\rm LF}}{0.94}\right)^2} &{(\small \cos\theta_{\rm LF}\geq 0.06)}.
  \end{array}
  \right.
  }
  \label{eq:bundle_cut}
\end{equation}
They are optimized for the cosmogenic neutrino model~\cite{Ahlers2010} 
with the least model rejection potential technique~\cite{mrf}.
Figure~\ref{fig:MuonBundleCut} shows the signal and background event distributions as a function of
$\cos\theta_{\rm LF}$ and $\log_{10}{{\rm NPE}}$. The muon bundle cut criteria
(Eq.~\ref{eq:bundle_cut}) are shown by the solid line in the figure.

The passing rates in each stage of the cuts with the IC86 configuration are described in Table~\ref{tbl:passing_rates}.
The expected number of atmospheric background events in the nine-year data sample
passing the selection criteria is $0.085$. The expected event rate
from a cosmogenic model~\cite{Ahlers2010} assuming the UHECR primaries to be dominated by protons
is 3.7 -- 7.0. This model~\cite{Ahlers2010} takes into account Fermi-LAT bounds on the $\gamma$-ray background 
generated by cascading of the high energy photons and electrons which are also produced in the GZK interactions. 
The range of the predicted cosmogenic neutrino flux (see TABLE~\ref{table:modelDependentLimit}) corresponds to 
different choices of the 'crossover' energy (1, 3 or 10 EeV) above which the extragalactic UHECR dominates over the Galactic component.
The astrophysical neutrino flux~\cite{IceCubeHESE2014}
can extend to the EHE region, and will yield an astrophysical background
with rates of $\lesssim 6$ events in
the nine-year analysis sample, depending on its spectral shape.

\section{\label{sec:likelihood} Binned Poisson likelihood method}
\begin{figure*}
  \includegraphics[width=0.4\textwidth]{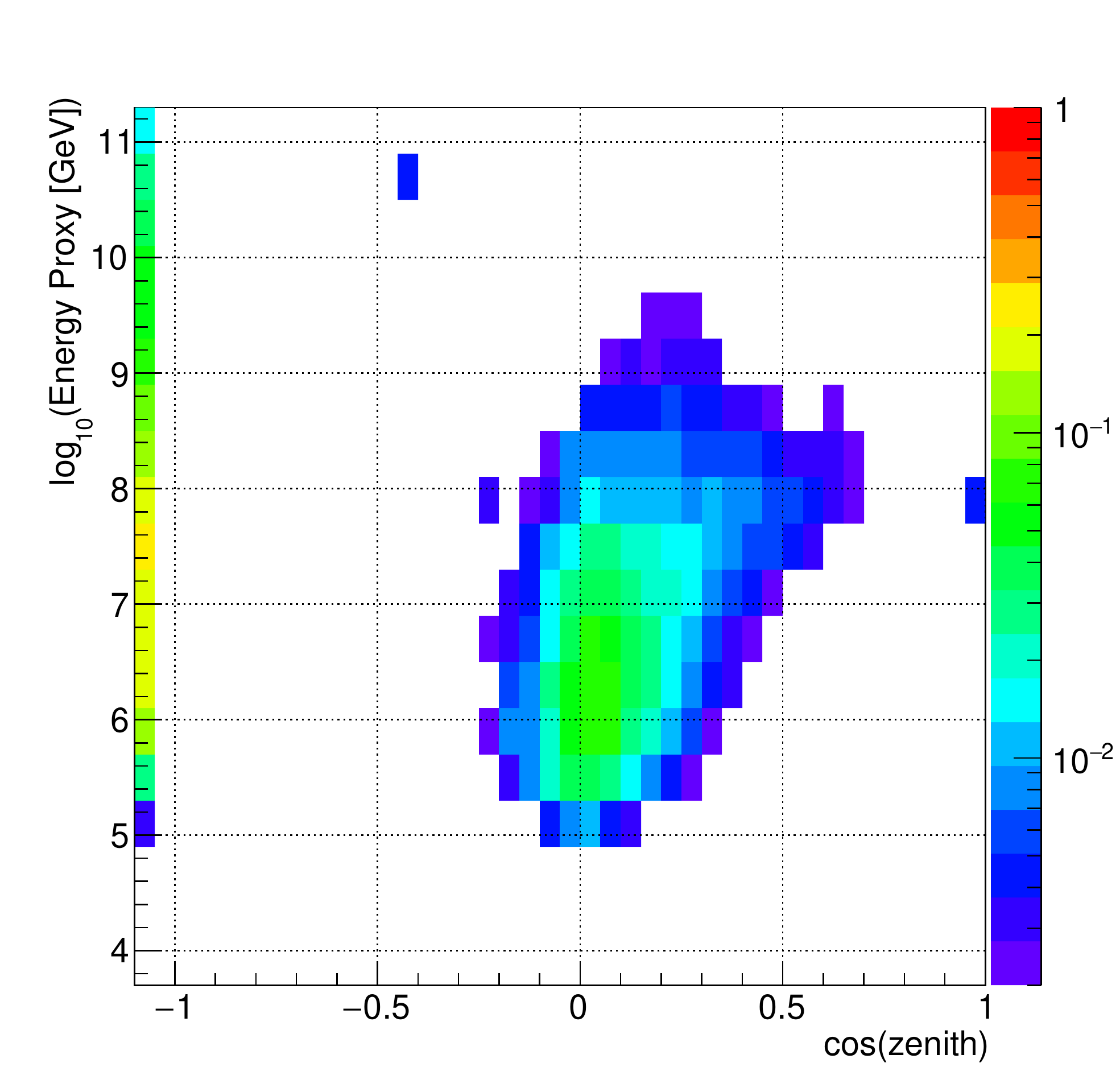}
  \includegraphics[width=0.4\textwidth]{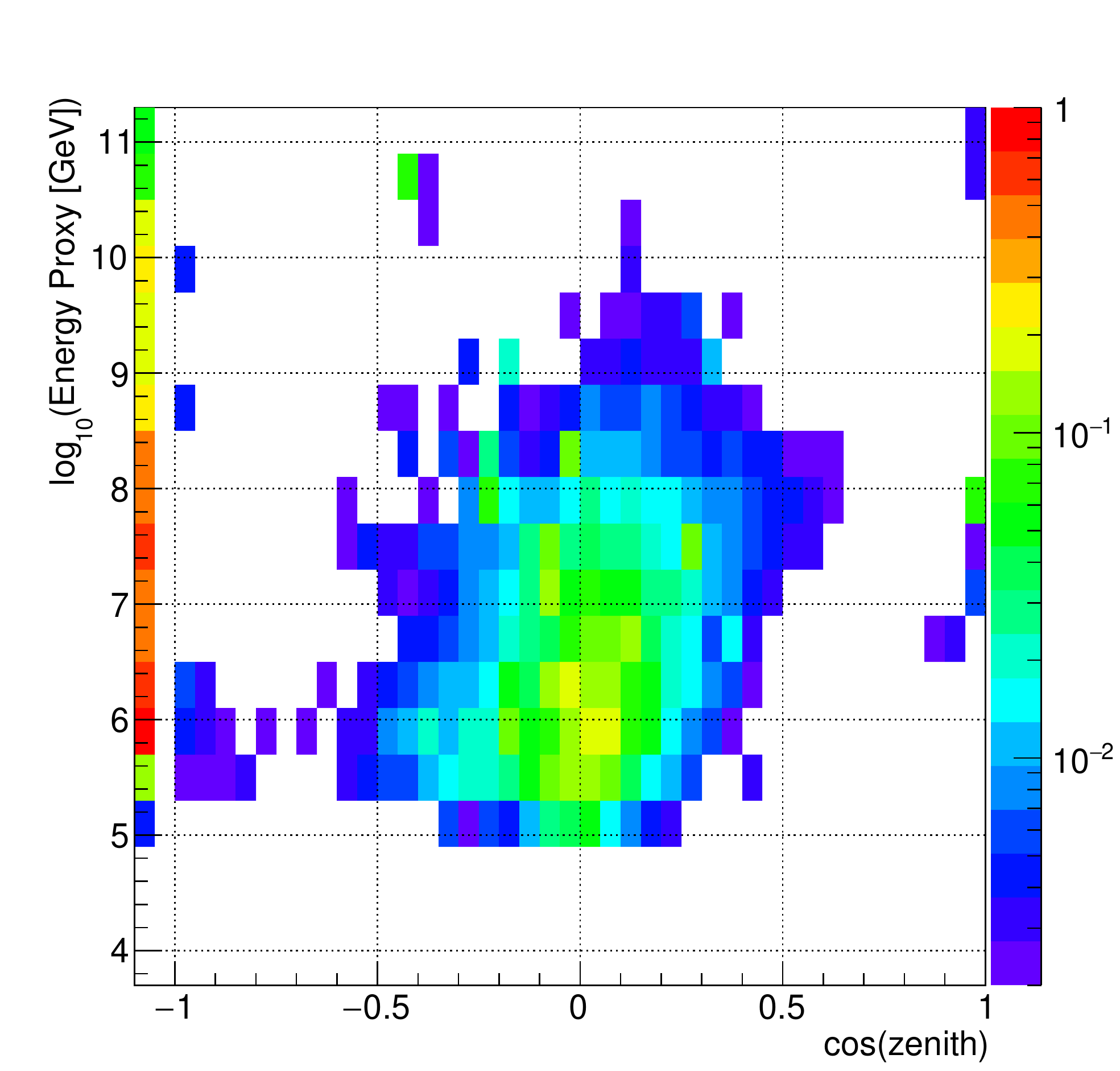}
  \caption{Event distributions as functions of the energy proxy
and cosine of the reconstructed zenith angle for simulations of the (left panel)
cosmogenic (GZK) model~\cite{Ahlers2010}
and (right panel) astrophysical neutrino model with a $E_\nu^{-2}$ spectrum
with an intensity of $E_\nu^2\phi_{\nu_e+\nu_\mu+\nu_\tau} = 10^{-8}\ {\rm GeV}/{\rm cm}^2\ \sec\ {\rm sr}$.
    The colors indicate the expected number of events seen by the IceCube
EHE neutrino analysis based on the data collected over nine years.
Rare misreconstructed events are
distributed in the unphysical region of the energy-direction parameter space and included in the figure.
 Events classified in the non-track-like category are plotted in the bins of $\cos({\rm zenith}) = -1.1$.}
  \label{fig:energy_cosZenith}
\end{figure*}

\subsection{\label{subsec:model_test} General model test}

In this analysis, observations are tested against theoretical models using a
binned Poisson likelihood method, which is defined as the product of the Poisson probabilities
over all zenith and energy bins as
\begin{equation}
L(\lambda) = \prod\limits_{i,j}P(n_{i,j};
\lambda\mu^{\rm SIG}_{i,j}+\mu^{\rm BG}_{i,j}),
\label{eq:Poisson_lh}
\end{equation}
where $P(n;\mu)$ is a Poisson PDF of observing $n$ events with the expectation of $\mu$ events.
$\mu^{\rm SIG}_{i,j}$ and $\mu^{\rm BG}_{i,j}$ are the mean number of the signal and background
(atmospheric neutrino and muons) events, respectively,
as functions of the cosine of the zenith angle (represented by bin $i$) and logarithm of an energy proxy defined below (bin $j$).
The data are binned in 42 zenith bins, and 32 energy proxy bins for this analysis.
The multiplier for a signal model, $\lambda$, 
can be varied in the test statistic construction.
$\lambda=1$ represents the predicted signal model strength.
Figure~\ref{fig:energy_cosZenith} presents some example event distributions.
The energy proxy used here is an energy deposition reconstruction that employs a
single-muon hypothesis with a series of stochastic energy losses from cascades along the muon track~\cite{reco}.
In the present analysis, the energy deposition reconstruction is 
specifically optimized to minimize the number of failed fits
so that no additional fit quality selections are required to obtain the event distributions.
The energy reconstruction method also shows reasonable performance for cascade-like events.
The resulting resolution of the energy proxy is approximately 0.8 decade
for through-going tracks, and 0.5 decade for contained cascade-like events.
Though the stochastic nature of EHE track energy loss profile broadens the resolution,
this energy deposition measure offers sufficient
correlations with neutrino energies to perform statistical tests on a given model flux.
The zenith angle, $\theta$, used here is provided by the single photoelectron log-likelihood fit~\cite{spe12}
based on the track hypothesis.
Events with the log-likelihood values inconsistent with the track hypothesis are categorized
in the {\it non-track-like} category.
Directional information for non-track-like events is not used in the analysis.

A model test is performed by comparing the model hypothesis
of $\lambda=1$ against
the alternative hypothesis $\lambda\neq 1$. The test statistic is the log-likelihood ratio:
\begin{equation}
  \Lambda = \log{\frac{L(\hat{\lambda})}{L(\lambda=1)}},
  \label{eq:llh_ratio}
\end{equation}
where $\hat{\lambda}$ is the multiplier that maximizes the Poisson likelihood $L$
by floating $\lambda$ between zero and infinity.
An ensemble of pseudo-experiments under the
model hypothesis is used to produce a PDF
of the test statistic $\Lambda$.
The p-value for a given model of cosmic neutrinos is then calculated
from the PDF by the frequency where $\Lambda$ is larger
than the $\Lambda$ observed.  A test of the atmospheric background only hypothesis is also conducted, where
$\lambda=0$.

\subsection{\label{subsec:model_comparison} Model comparison}

The binned Poisson likelihood introduced for the cosmogenic (GZK) model ($L_{\rm GZK}$) and
power-law model ($L_{\rm power}$) can be written as
\begin{eqnarray}
L_{\rm GZK}(\lambda_{\rm GZK}) &=& \prod\limits_{i,j}P(n_{i,j};
  \lambda_{\rm GZK} \mu^{\rm GZK}_{i,j}+\mu^{\rm BG}_{i,j}),\nonumber\\
L_{\rm power}(\lambda_{\alpha}) &=& \prod\limits_{i,j}P(n_{i,j};
  \lambda_{\alpha} \mu^{\alpha}_{i,j}+\mu^{\rm BG}_{i,j}),
  \label{eq:Poisson_llh_model}
\end{eqnarray}
where $\mu^{\rm GZK}_{i,j}$ is the number of events in a bin of the energy--zenith plane
predicted by the cosmogenic model and $\mu^{\alpha}_{i,j}$ is the value attributable to
a generic astrophysical $E_\nu^{-\alpha}$ power-law flux.
One important question is whether the observed data are consistent
with the expectations from cosmogenic neutrino models~\cite{berezinsky69}
or a softer power-law
flux, such as $E_\nu^{-2}$, as expected from astrophysical neutrinos.
The test statistic here is
\begin{equation}
  \Lambda = \log{\frac{{L_{\rm power}}(\widehat{\lambda_{\alpha}})}{L_{\rm GZK}(\widehat{\lambda_{\rm GZK}})}},
  \label{eq:llh_ratio_gzk_astro}
\end{equation}
where $\widehat{\lambda_{\alpha}}$ and $\widehat{\lambda_{\rm GZK}}$ maximize the likelihood functions.

\subsection{\label{subsec:astro_nuisance} Calculations with astrophysical background}

The astrophysical neutrino flux observed by IceCube indicates that contributions
from a generic astrophysical power-law flux are expected in the PeV energy region~\cite{EHE2016}.
We account for this possibility
by introducing a nuisance flux in the form $\phi_\alpha=\kappa_\alpha E_\nu^{-\alpha}$,
where $\kappa_\alpha$ is an arbitrarily chosen reference normalization.
A small modification of Eq.~(\ref{eq:Poisson_llh_model}) gives
\begin{equation}
L_{\rm GZK}(\lambda_{\rm GZK},\lambda_\alpha) = \prod\limits_{i,j} P(n_{i,j};
\lambda_{\rm GZK} \mu^{\rm GZK}_{i,j}+\lambda_\alpha\mu^\alpha_{i,j}+\mu^{\rm BG}_{i,j}).
\label{eq:llh_gzk}
\end{equation}
Taking $\lambda_\alpha$ as a nuisance parameter, the likelihood ratio
is constructed using the profile likelihood:
\begin{equation}
  \Lambda(\lambda_{\rm GZK}) = \log{\frac{L_{\rm GZK}(\widehat{\lambda_{\rm GZK}},\widehat{\lambda_\alpha})}
    {L_{\rm GZK}(\lambda_{\rm GZK},\widehat{\widehat{\lambda_\alpha}}(\lambda_{\rm GZK}))}},
  \label{eq:llh_ratio_gzk}
\end{equation}
where the double-hat notation represents the profiled value of the parameter
$\lambda_\alpha$, defined as the value that maximizes $L_{\rm GZK}$ for the specified $\lambda_{\rm GZK}$.
This likelihood ratio, in which $\lambda_{\rm GZK}=1$,
is the test statistic for a given cosmogenic neutrino model.
The baseline model of the nuisance flux is built with $\alpha=2$.
The impact of different power-law indices is negligible when constraints are placed in the EHE region
because we confirmed that upper limits of $\lambda_{\rm GZK}$ with various $\alpha$ ranging from 2.5 to 2.0 are completely consistent
within the statistical precision of pseudo-experiments to produce a PDF of $\Lambda$.
The recent model-dependent p-values and the upper limits
for the selected cosmogenic models in~\cite{EHE2016}
were obtained using this procedure.

\subsection{\label{subsec:dif_limit_cal} Extension to differential limit}

\begin{figure*}[t]
  \includegraphics[width=0.32\textwidth]{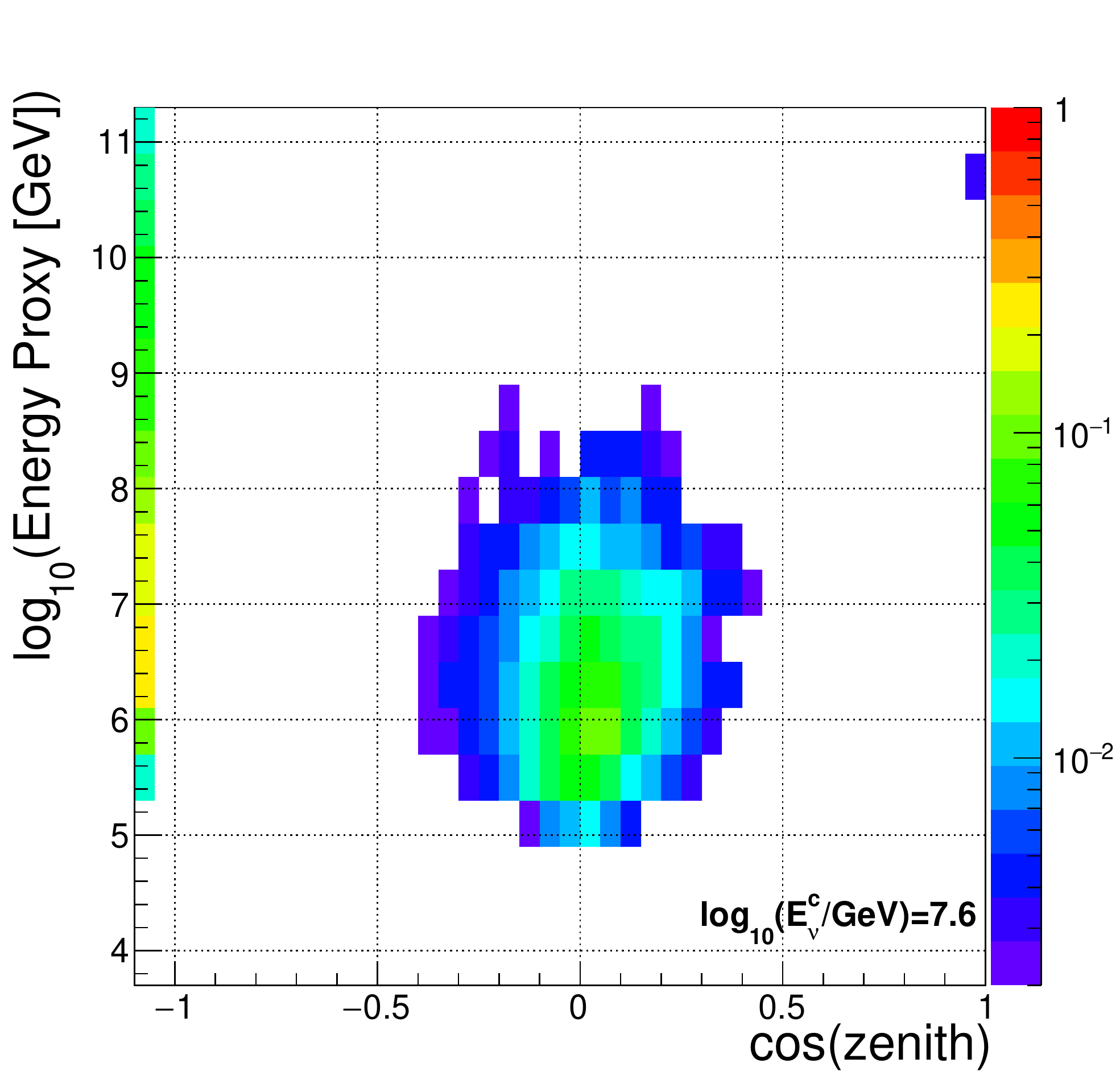}
  \includegraphics[width=0.32\textwidth]{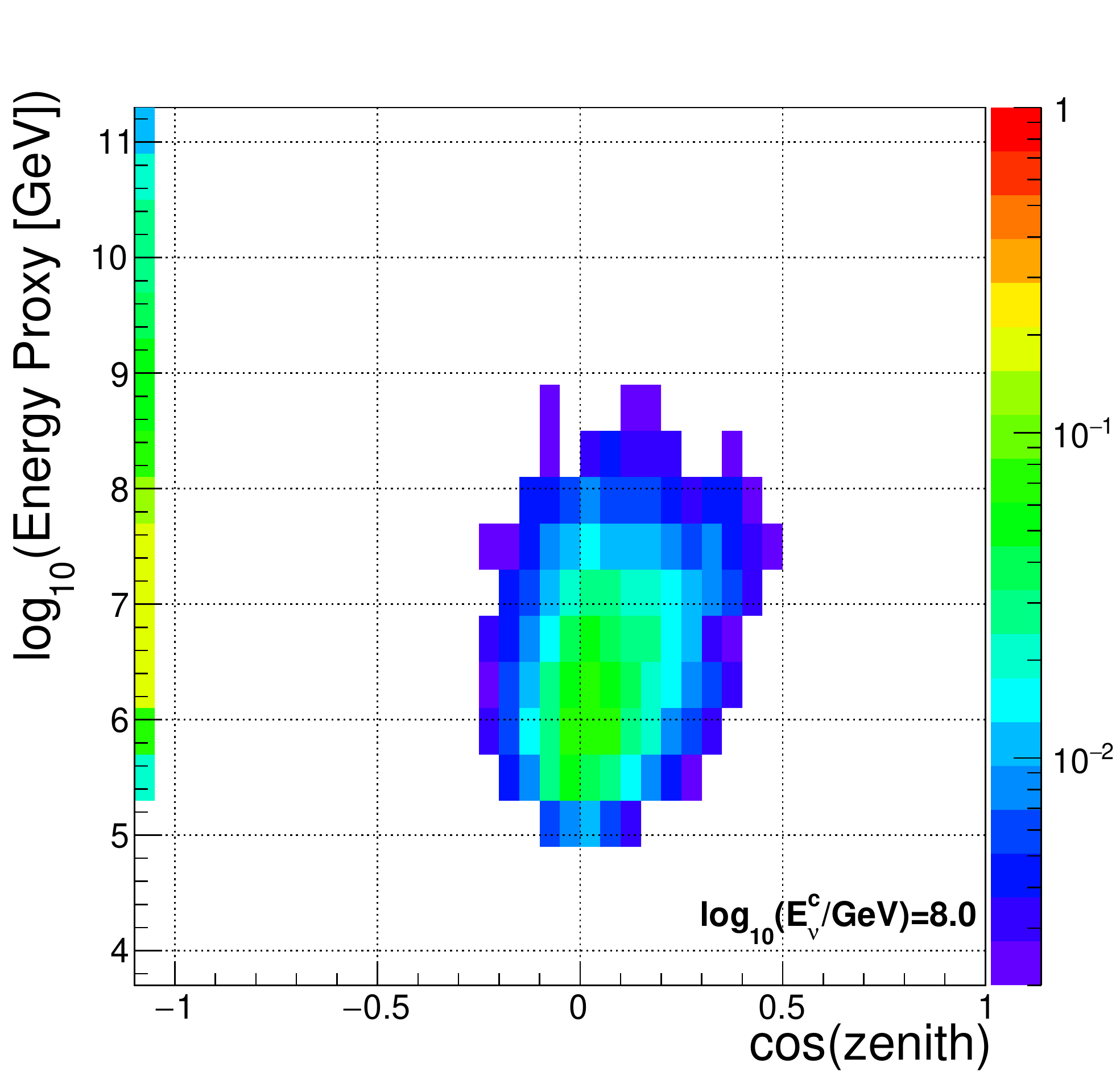}
  \includegraphics[width=0.32\textwidth]{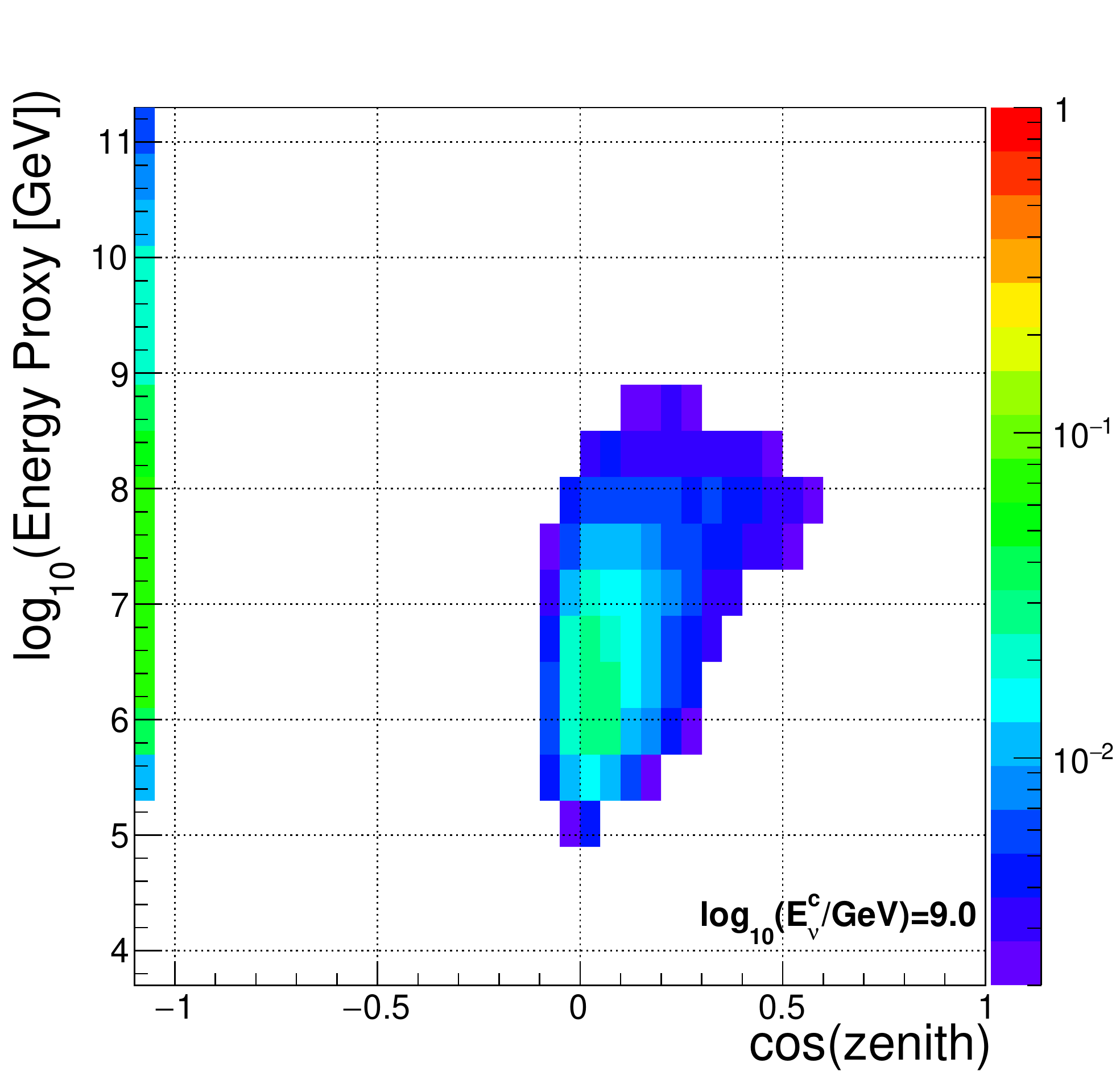}
  \caption{Event distributions as functions of the energy proxy
and cosine of the reconstructed zenith angle for the flux
$\phi_{\rm diff}=\kappa_E E_\nu^{-1}$, spanning a
one-decade energy interval centered at $E_\nu^c$. The event distributions include the contributions
from all three neutrino flavors.
Events classified as non-track-like
are plotted in the bins of $\cos({\rm zenith}) = -1.1$. From left to right,
the distributions for $\log_{10}{(E_\nu^c/{\rm GeV})}=7.6, 8.0$,
and $9.0$ are shown.
Note that this energy proxy was designed to work across all event topologies for the EHE analysis.
Better energy estimates are obtained by dedicated energy reconstructions 
optimized for a specific event topology.
For display purposes, the normalization $\kappa_E$ has been set here
so the energy flux
$E_\nu^2\phi_{\rm diff} = 1.0\times 10^{-8}\ {\rm GeV}/{\rm cm}^2\ \sec\ {\rm sr}$
at an energy of $E_\nu^c$.}
  \label{fig:energy_cosZenith_E1}
\end{figure*}

The inclusion of an astrophysical nuisance parameter can be extended to the differential limit calculation.
The differential limit at a neutrino energy of $E_\nu^c$ presented here is the limit
for the flux of $\phi_{\rm diff} = \kappa_E E_\nu^{-1}$ ranging over an interval of one decade
$[\log_{10}(E_\nu^c/{\rm GeV})-0.5,\log_{10}(E_\nu^c/{\rm GeV})+0.5]$.
A generalized hypothesis test in the presence of an astrophysical flux can similarly
be obtained with Eq.~(\ref{eq:llh_gzk}).
Here, instead of using a cosmogenic model flux, $\phi_{\rm diff}$ is used. Thus,
\begin{equation}
L_{\rm diff}(\lambda_{\rm diff},\lambda_\alpha) = \prod\limits_{i,j} P(n_{i,j};
\lambda_{\rm diff} \mu^{\rm diff}_{i,j}(E_\nu^c)+\lambda_\alpha\mu^\alpha_{i,j}+\mu^{\rm BG}_{i,j}),
\label{eq:llh_dif}
\end{equation}
where $\mu^{\rm diff}$ represents contributions from the flux $\phi_{\rm diff}$ with a one-decade
energy interval centered at $E_\nu^c$. Thus, this expression is a function of $E_\nu^c$.
Figure~\ref{fig:energy_cosZenith_E1} presents the
distribution of $\mu^{\rm diff}_{i,j}$
in the energy--zenith angle plane.
The differences in the energy proxy between various $E_\nu^c$ are not substantial, because the
deposited energy of a secondary muon track is only weakly correlated to the primary neutrino energy.
This quality of resolution arises from the stochastic nature of the muon-energy-loss profile
at PeV--EeV energies, the large variance in the fraction of neutrino energy
channeling into muons, as well as variations in the position where muons are created.
Instead, the zenith angle distribution exhibits more $E_\nu^c$ dependence.
The larger the value of $E_\nu^c$, the more events are distributed
above the horizon, where $\cos({\theta})\geq0$.
This occurs because neutrinos with higher energies experience stronger absorption effects
during their propagation through the Earth. The zenith angle distribution is
a key feature for setting the differential limit at energies higher than $10^{7}$ GeV.

The test statistic is constructed as
\begin{equation}
  \Lambda(\lambda_{\rm diff},E_\nu^c) =
  \log{\frac{L_{\rm diff}(\widehat{\lambda_{\rm diff}(E_\nu^c)},\widehat{\lambda_\alpha})}
    {L_{\rm diff}(\lambda_{\rm diff}(E_\nu^c),\widehat{\widehat{\lambda_\alpha}}(\lambda_{\rm diff}(E_\nu^c)))}}.
  \label{eq:llh_ratio_dif_limit}
\end{equation}

\begin{figure*}
  \includegraphics[width=0.32\textwidth]{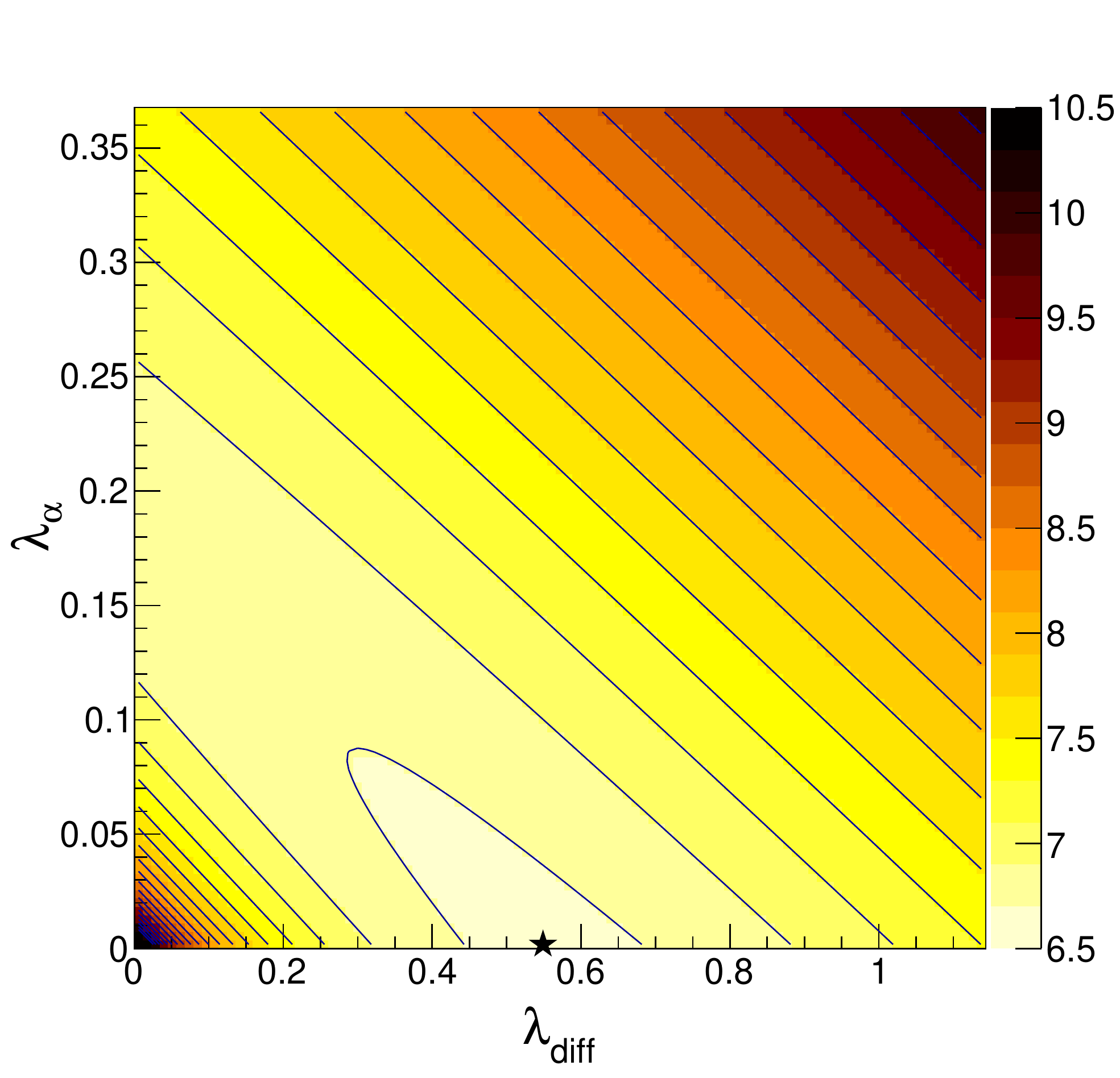}
  \includegraphics[width=0.32\textwidth]{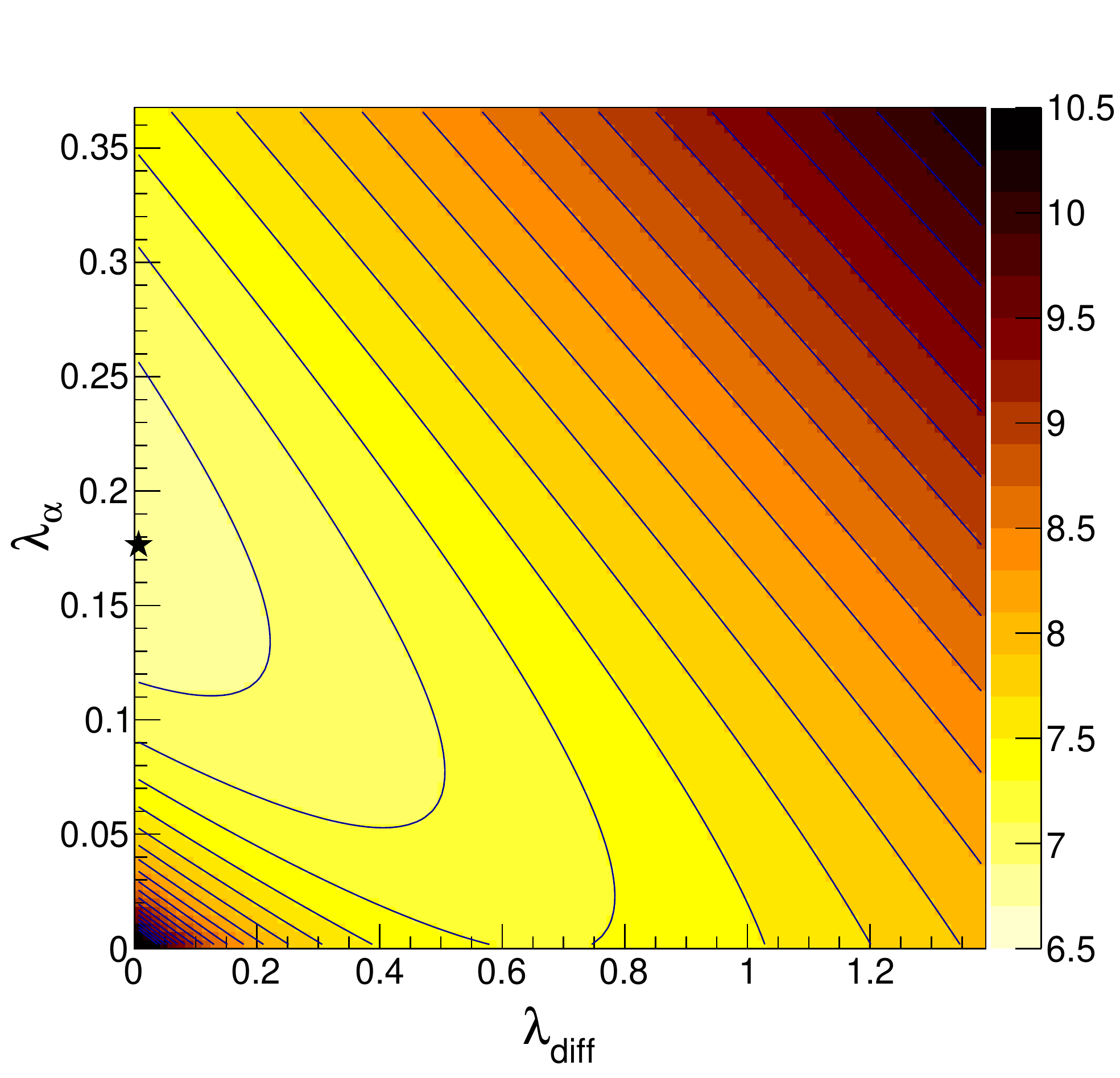}
  \includegraphics[width=0.32\textwidth]{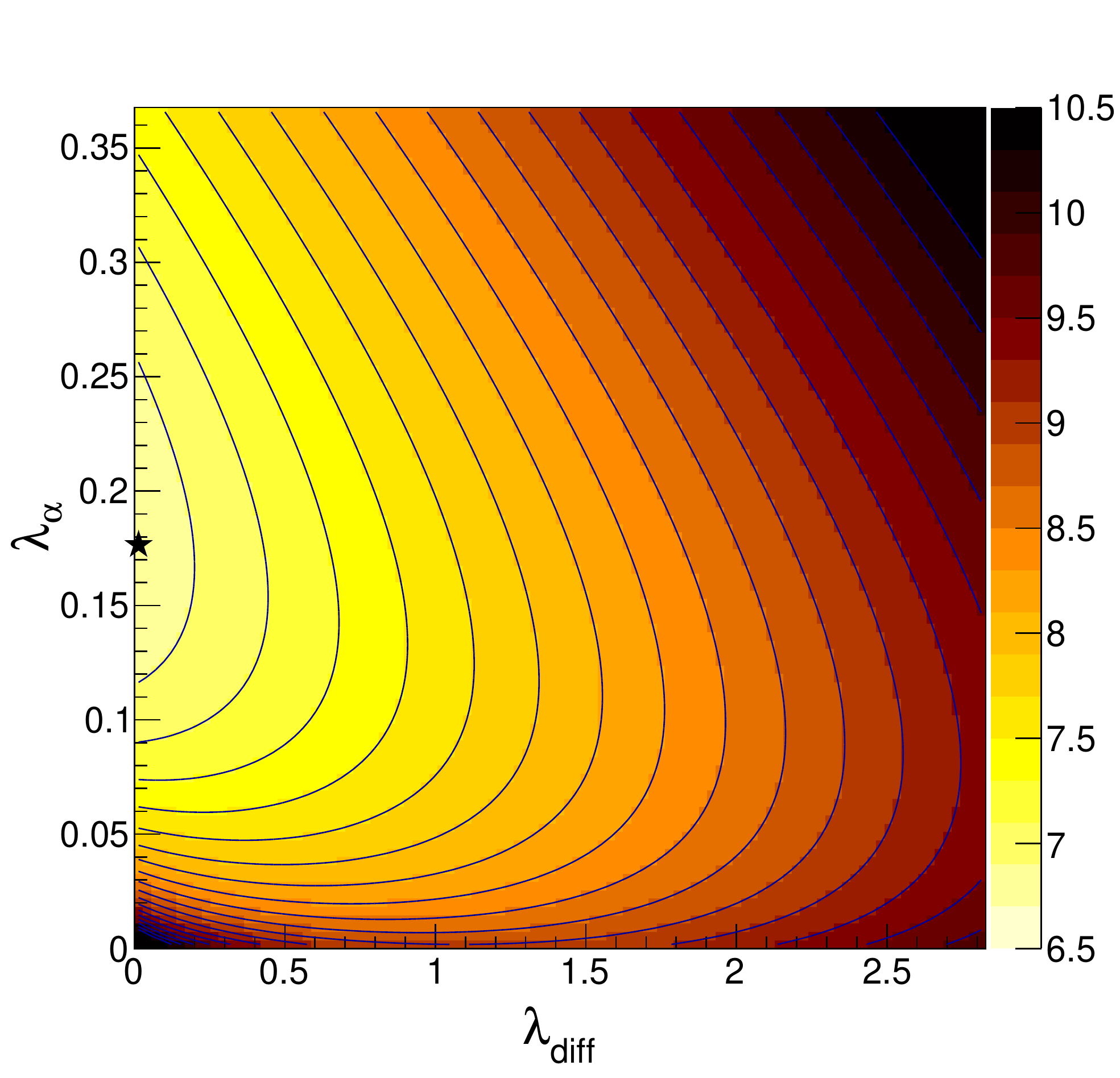}
  \caption{Distribution of the negative log-likelihood ($-\log{L_{\rm diff}}$) for the hypothesis of
    $E_\nu^{-1}$ flux ranging over an energy interval of one decade,  centered at $E_\nu^c$, on the
    $\lambda_{\rm diff}$-$\lambda_{\alpha}$ plane. Here $\lambda_{\rm diff}$ denotes the multiplier
    to the one-decade box-type spectrum $\phi_{\rm diff}$ and $\lambda_{\alpha}$ is the multiplier
    to the power-law nuisance spectrum $\phi_{\alpha}$ representing the astrophysical background.
    The star indicates the minimum on the $-\log{L_{\rm diff}}$ plane.
    From left to right, the distributions for
    $\log_{10}{(E_\nu^c/{\rm GeV})}=7.6, 8.0$, and $9.0$ are shown.
    These results are calculated using the nine-year set of IceCube data containing
    two PeV-energy events described in Sec.~\ref{sec:results}. The power-law index
    of astrophysical flux, $\alpha$, is set to 2 in these examples.}
  \label{fig:llh_scan}
\end{figure*}

An ensemble of pseudo-experiments is used to construct the PDF of $\Lambda(\lambda_{\rm diff},E_\nu^c)$,
and gives the upper limit of $\lambda_{\rm diff}$ at a given confidence level,
for an energy of $E_\nu^c$. By repeating the same procedure with varying $E_\nu^c$,
the differential upper limit as a function of neutrino energy is produced.

Figure~\ref{fig:llh_scan} presents the distributions of the values
of the negative log-likelihood $-\log{L_{\rm diff}}$
for several values of the neutrino energy $E_\nu^c$. The local minimum point
on each of the $\lambda_{\rm diff}$-$\lambda_{\alpha}$ planes
corresponds to $\widehat{\lambda_{\rm diff}(E_\nu^c)},\widehat{\lambda_\alpha}$
that maximize the likelihood.
At $E_\nu^c = 10^{7.6}\ {\rm GeV}$, the minimal point is found at
$\lambda_\alpha=0$, which implies that the observational data including
the two detected events (see Sec~\ref{sec:results})
are attributed to $\phi_{\rm diff}$ centered
at an energy of $10^{7.6}$ GeV, and do not require an astrophysical nuisance flux.
This result occurs because the primary energies of the neutrinos initiating the detected events
are likely to originate in the one-decade-energy interval of $\phi_{\rm diff}$.
In the case of the central energy $E_\nu^c$ of $10^{8}\ {\rm GeV}$,
$\lambda_{\rm diff}=0$ maximizes the likelihood, implying that the data
disfavor the one-decade box-type spectrum $\phi_{\rm diff}$
but prefer a nonzero component of the astrophysical nuisance flux.
Further increases in the central energy $E_\nu^c$ weaken the correlation between $\lambda_\alpha$ and
$\lambda_{\rm diff}$
and the upper limit of the one-decade box flux $\phi_{\rm diff}$ becomes less dependent
on the intensity of the astrophysical nuisance flux, as one can see
in the far-right plot of Fig.~\ref{fig:llh_scan}. Because the differential limit
corresponds to the 90\% CL~upper limit of $\lambda_{\rm diff}$, this method gives the limit
on $\phi_{\rm diff}$  in the presence of a possible astrophysical flux
whose intensity $\lambda_\alpha$ is estimated by
the real data sample.
The estimated astrophysical neutrino intensity with $\alpha=2$ is
$E_\nu^2\phi_{\nu_e+\nu_\mu+\nu_\tau} = 1.7\times 10^{-9}\ {\rm GeV}/{\rm cm}^2\ \sec\ {\rm sr}$
regardless of $E_\nu^c$ when $E_\nu^c \gtrsim 5\times 10^7$ GeV.
The obtained limit is robust against the different assumptions about the astrophysical neutrino
spectrum, such as a softer spectrum of $\alpha=2.5$ or a spectral cutoff at 3 PeV
as the resultant limit changes only by $\alt 5\%$ with these various spectral assumptions.

For the previously published differential EHE limit~\cite{EHE2016},
no nuisance parameter was used to account for an astrophysical
background neutrino flux and the limit applies to the total neutrino
flux over a decade in energy. Employing the astrophysical nuisance parameter
in calculation of the differential limit, 
one must consider that
the PDF of the test statistic $\Lambda$,
given by Eq.(\ref{eq:llh_ratio_dif_limit}), depends on 
the true value of the multiplier for the astrophysical flux
in contrast to the consequence of the widely used Wilks' theorem 
for high statistical data samples. It implies that the resultant limit may depend
on the nuisance flux multiplier $\lambda_\alpha$, whose true value is yet to be understood.
For setting the differential limit, we calculated the PDF of the test statistic $\Lambda$ 
by pseudo-experiments with various astrophysical flux intensities and found that 
the case of no astrophysical background resulted in the most conservative limit.
Though the likelihood function and test statistic do include
the nuisance astrophysical flux as a floating parameter (see Eq.~(\ref{eq:llh_ratio_dif_limit})), 
the differential limit presented here is, hence, derived by the $\Lambda$ distribution 
assuming no astrophysical flux.

\section{\label{sec:results} Results and discussion}

\begin{table*}
\begin{center}
  \caption{Characteristics of the detected events found in this analysis.
    The energy proxy values listed here represent the estimates of energy deposition
    that are used for building the binned Poisson likelihood in the present analysis.
    They are obtained by the event reconstruction designed to be applicable to
    the EHE event sample regardless of their event topology.
    The best-estimated $\nu$ energy displays the parent neutrino energy estimates
    obtained by dedicated event reconstructions optimized for each event topology.}
\label{tb:events}
{\small
\begin{tabular}{cc@{\hspace{0.8cm}}c@{\hspace{0.8cm}}c}\hline \hline
    & Energy proxy in the present analysis [PeV] & Best estimated $\nu$ energy [PeV]& Event topology \\ \hline
event 1 & 2.6 & 8.7 (median~\cite{diffuseNuMu6yr}) & track   \\
event 2 & 2.7 & 5.9  & uncontained shower \\
\hline \hline
\end{tabular}
}

\end{center}
\end{table*}

Two events passing the final selection criteria were observed;
one event was reported in the previous analyses~\cite{EHE2016,diffuseNuMu6yr},
and the newly found event in the additional two-year sample was detected in December 2016.
It appears as a partially contained shower event.
The energy proxy of this event used in the present analysis ($E_{proxy}$) is 2.7~PeV.
Note that the best-estimated energy of this uncontained shower event
is different from the energy proxy value. A dedicated energy loss reconstruction algorithm
based on extensive simulations of this type of event estimate its energy as 5.9~PeV.
Additional details will be published elsewhere.
The characteristics of the observed events are listed in Table~\ref{tb:events}.

The hypothesis that these two events are backgrounds
of atmospheric origin
was tested by the likelihood ratio test statistic of Eq.~(\ref{eq:llh_ratio}) with $\lambda=0$
and is rejected with a p-value of 0.024\% ($3.5\sigma$). They are found compatible with
a generic astrophysical $E^{-2}$ power-law flux with a p-value of 78.8\%,
whereas they are inconsistent with the cosmogenic hypothesis
with a p-value of 2.5\% ($2.0\sigma$), calculated using the test statistic of Eq.~(\ref{eq:llh_ratio_gzk_astro})
employing the GZK neutrino model by Ahlers {\it et al.}~\cite{Ahlers2010}.
The two observed events are more consistent with neutrinos from astrophysical power-law flux extending from
TeV to PeV energies than from the cosmogenic flux peaking at energies in the EeV range.

The systematic uncertainties are the same as in the previous analysis~\cite{EHE2016}
and each of the sources of systematic errors is fully described in Ref.~\cite{EHE2012}.
The upper limits are weakened primarily by a potential NPE shift
due to uncertainties in the detector's optical detection efficiency, and potential
signal reduction due to uncertainties in the neutrino--nucleon cross section.
Differential limits are derived including the worst-case combinations of these uncertainties.
The effective softening of the limit was by about $28\%$ below $4\times 10^8$~GeV
and by about $11\%$ at about $10^9$ GeV and above.

\begin{figure}[t]
\begin{center}
  \includegraphics[width=0.5\textwidth]{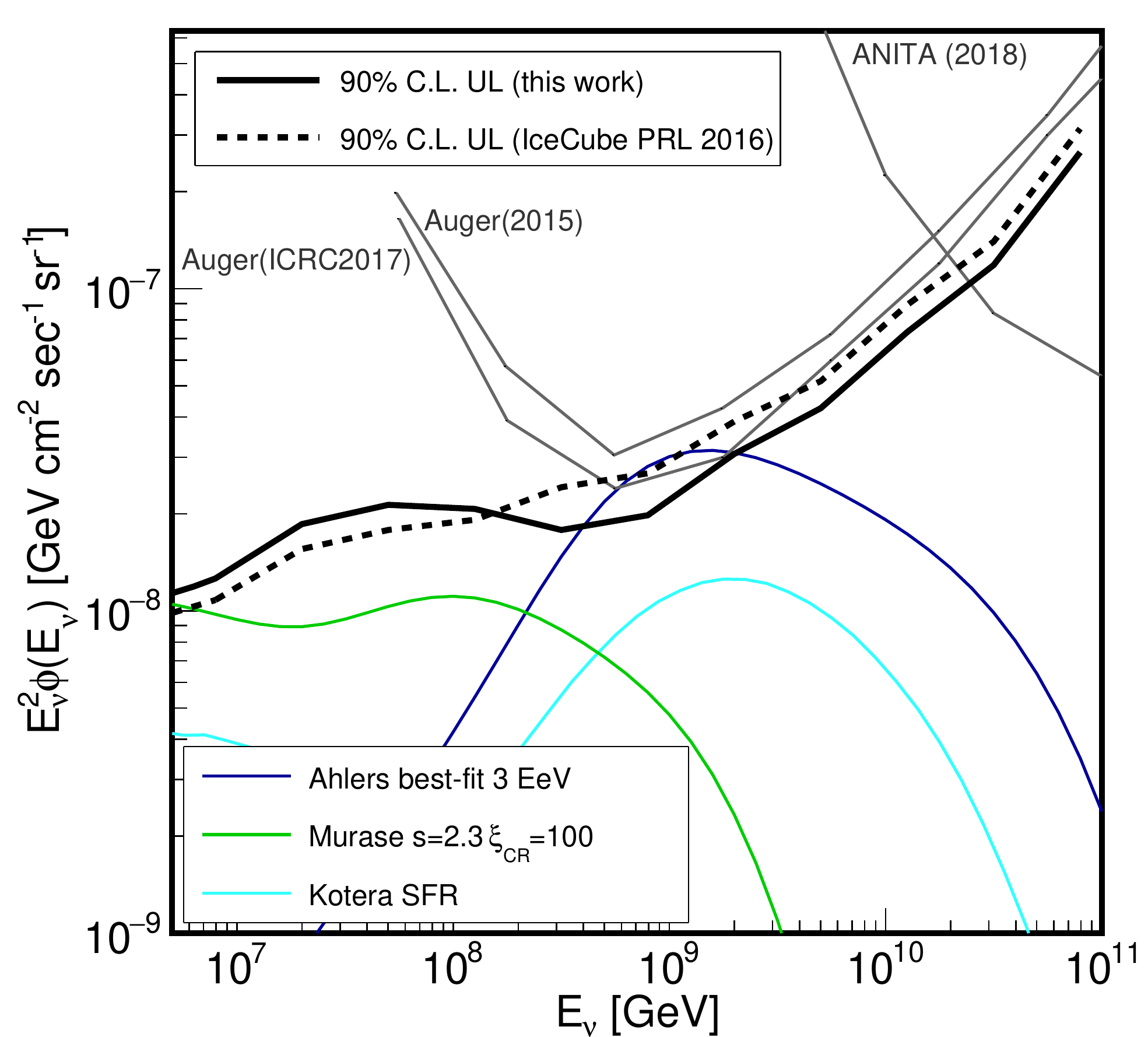}
  \caption{All-flavor differential 90\% C.L. upper limit
           based on the nine-year sample of IceCube data (solid line).
           Cosmogenic neutrino model predictions (assuming primary protons)
           by Kotera {\it et al.}~\cite{Kotera2010} and Ahlers {\it et al.}~\cite{Ahlers2010},
           and an astrophysical neutrino model by Murase {\it et al.}~\cite{murase2014} are
           shown for comparison. Differential limits
           for one-energy-decade $E_\nu^{-1}$ flux by other experiments are also shown for
           Auger (2015)~\cite{augerNu2015}, (ICRC2017)~\cite{augerICRC2017}, and ANITA~\cite{anita}
           with appropriate normalization by considering the energy bin width
           and neutrino flavor. The previous IceCube limit from the analysis
           of seven years of data~\cite{EHE2016} with the similar likelihood ratio framework but
           without a nuisance astrophysical background flux parameter is
           also shown for reference (dashed line).
}
  \label{fig:DifLimit}
\end{center}
\end{figure}
Figure~\ref{fig:DifLimit} presents the differential upper limit
on the all-flavor neutrino flux using
this new method based on the nine-year sample of IceCube data.
The two observed events weaken the limit below $4\times 10^8\ {\rm GeV}$,
while the limit becomes more stringent at higher energies 
as the astrophysical background
completely accounts for the detected events. In the energy range most relevant to UHECR emissions,
the present limit is stronger than the previous IceCube limit~\cite{EHE2016}
even though the number of events remaining in the final data sample has doubled from one event to two.
The new method for calculating differential upper limits with the nuisance flux strengthens the limit by $\sim 45$ \%
in the energy region around $10^9$ GeV in addition to the statistical improvements
by adding two years of data.
The limit applies to the constraints of the EHE cosmic neutrino flux on top of
a power-law flux of astrophysical neutrinos.
Any departure from $\alpha=2$ in the nuisance $\phi_\alpha$ model has a very minimal impact
on the obtained limit, especially at energies of $3\times 10^{8}$ GeV or higher, the main energy
region of interest for this study. The presented limit is also insensitive to
systematic uncertainties in the energy proxy and topology of the detected events.

The presented differential upper limit in the energy region
between $5\times 10^6$ and $2\times 10^{10}$ GeV is the most constraining model-independent
upper limit currently reported. Models predicting a flux of
$E_\nu^2\phi_{\nu_e+\nu_\mu+\nu_\tau}\simeq2\times 10^{-8}\ {\rm GeV}/{\rm cm}^2\ \sec\ {\rm sr}$
at $10^9\ {\rm GeV}$ are disfavored by the IceCube observations.
Although the newly detected PeV-energy neutrino event relaxed the present limit
below $4\times 10^8\ {\rm GeV}$, the obtained differential limit
represents our most recent model-independent bound given by large IceCube exposure.

The present limit constrains a significant portion of the parameter space
in EHE neutrino models that assume a proton-dominated UHECR composition.
This constraint arises because the energy flux of UHECRs at 10 EeV,
about $2\times 10^{-8}\ {\rm GeV}/{\rm cm}^2\ \sec\ {\rm sr}$, is comparable
to the present neutrino differential limit.
The UHECR flux is contributed only from
sources in the local universe within a distance of $R_{\rm GZK}\sim 100\ {\rm Mpc}$
because of the energy attenuation of UHECR protons colliding with the cosmic microwave background.
However, neutrinos are able to travel cosmological distances of $O(c/H_0)\sim 4\ {\rm Gpc}$.
Thus, UHECR sources within a sphere of about $c/H_0$ contribute to the expected neutrino flux.
This volume effect generally increases the neutrino flux relative to
the UHECR flux by a factor of about $c/H_0/R_{\rm GZK}\sim O(10)$.
This balances the energy conversion factor
from a UHECR proton to its daughter neutrino ($5-10\%$),
leading to an amount of neutrino energy flux comparable
to the energy flux of UHECRs, if the observed UHECRs are protons,
independent of the details of the neutrino production model.
The present improved limit above $10^8$~GeV on the proton-dominated UHECR composition model is, therefore,
robust against theoretical and observational uncertainties such as
the cosmogenic neutrino intensity at PeV energies, which is determined by 
the extragalactic background light whose intensity is still uncertain~\cite{EBL2017}.
Constrained by the differential limit, the interaction model-independent 
constraints~\cite{Heinze2016} can be applied to
the UHECR transition/composition models such as the proton dip model~\cite{protonDip}.

\begin{table}
  \caption{Neutrino model tests: Expected number of events,
    p-values from model hypothesis test, and 90\% C.L. model-dependent limits in terms of
    the model rejection factor (MRF)~\cite{mrf}, defined as the ratio
    between the flux upper limit and the predicted flux.
    The systematic uncertainties are taken into account in
    setting the MRF's and the errors on p-values.
    All of the models listed here
    assume proton-dominated UHECRs, except for the AGN model by Murase {\it et al}
    where the baryon loading factor $\xi_{\rm CR} =100$ so that the injected cosmic rays
    from the blazars can  achieve the observed UHECR generation rate around $10^{10}-10^{10.5}$~GeV.
    }
\label{table:modelDependentLimit}
\begin{center}
\begin{tabular}{lccc}
\hline
\hline
$\nu$ Model & Event rate & p-value & MRF\\\hline
\hline
Kotera {\it et al.}~\cite{Kotera2010}   &  &  & \\
SFR         & 4.8    & 13.3$^{+6.5}_{-2.3}$\%   & 1.23 \\
Ahlers {\it et al.}~\cite{Ahlers2010}   &  &    & \\
best fit, 1 EeV &3.7  & 19.2$^{+13.2}_{-3.2}$\%  & 1.51 \\
Ahlers {\it et al.}~\cite{Ahlers2010}   &  &    & \\
best fit, 3 EeV & 5.7 &4.6$^{+2.6}_{-2.0}$\%  &  0.68 \\
Ahlers {\it et al.}~\cite{Ahlers2010}   &  &    & \\
best fit, 10 EeV &7.0  & 1.8$^{+4.1}_{-0.5}$\% & 0.63 \\
Aloisio {\it et al.}~\cite{Aloisio2015} &   &  & \\
SFR       & 6.3   & 4.1$^{+7.7}_{-0.9}$\%   & 0.93 \\
Murase {\it et al.}~\cite{murase2014} &   &  & \\
AGN, $s=2.3$, $\xi_{\rm CR}=100$  & 9.9   & 1.5$^{+6.7}_{-1.0}$\%   & 0.74 \\
\hline
\hline
\end{tabular}
\end{center}
\end{table}

While the differential upper limits provide a good indicator
of how the bound of EHE neutrino flux constrains different models,
the model-{\it dependent}
upper limits are more stringent in constraining each model.
This arises because the EHE neutrino models, in general, predict
neutrino fluxes ranging across several decades of neutrino energy.
This behavior is demonstrated by
the fact that the cosmogenic neutrino flux reported by Kotera {\it et al.}~\cite{Kotera2010}
and the active galactic nuclei (AGN) neutrino flux reported by Murase {\it et al.}~\cite{murase2014}, as
shown in Fig.~\ref{fig:DifLimit}, were disfavored in this analysis.
Table~\ref{table:modelDependentLimit} presents the results of the model-dependent tests
for selected cosmogenic models and
an astrophysical AGN model.
These constraints were obtained by following
the procedure described in Sec.~\ref{subsec:astro_nuisance} and are compatible with the analysis based on the
seven-year set of IceCube data~\cite{EHE2016}
though the newly detected PeV-energy event slightly weakens the constraints.

\section{\label{sec:summary} Summary}
In this study, an EHE neutrino search using a nine-year IceCube data set was conducted,
and we identified two distinct events with energies beyond 1 PeV. No events in
the energy region above 10 PeV were found. This observation indicates that
no neutrinos were induced by UHECR nucleons via the GZK mechanisms. This is consistent
with the model-dependent constraints previously published~\cite{EHE2016}
based on the seven years of data. It can be concluded that the cosmological evolution of UHECR
sources must be comparable to or weaker than the star formation rate,
a generic measure of structure formation history in the Universe~\cite{SFR},
if the mass composition of UHECRs is proton-dominated.
This finding is also consistent with the constraints from the diffuse extragalactic
$\gamma$-ray background~\cite{BrezinskyFermi,globus2017} measured by Fermi-LAT~\cite{fermi}.

In order to place an EHE neutrino flux limit with the present IceCube data set
containing astrophysical neutrino background  events,
we introduced a new method that employs a binned Poisson likelihood method
with a nuisance parameter to represent the TeV--PeV energy astrophysical neutrino flux.
The intensity of the nuisance flux is determined from the observed data
using a profile likelihood construction.
The obtained differential limit is the most stringent recorded to date in the energy range
between $5 \times 10^6$ and $2 \times 10^{10}$ GeV.
This indicates that any cosmic neutrino model
predicting a three-flavor neutrino flux of
$E_\nu^2\phi_{\nu_e+\nu_\mu+\nu_\tau}\simeq2\times 10^{-8}\ {\rm GeV}/{\rm cm}^2\ \sec\ {\rm sr}$
at $10^9\ {\rm GeV}$ is severely constrained. This is a universal bound of EHE cosmic neutrinos,
regardless of the model of the EHE neutrino production and their sources.

The present limits with IceCube observations significantly challenge
the most popular candidates for UHECR sources, such as $\gamma$-ray bursts and
radio-loud AGNs, but if the highest-energy cosmic rays are not proton-dominated,
it is clear that these constraints are weakened. A mixed-composition scenario,
in general, predicts EHE neutrinos with an intensity lower than the present bound
by an order of magnitude~\cite{globus2017}. A larger-scale neutrino detector
is required to measure the EHE neutrino flux in this case. Experimental
constraints on sources of UHECRs of mixed or heavy composition will be
provided in a next-generation detector such as ARA~\cite{ARA}, ARIANNA~\cite{ARIANNA},
or IceCube-Gen2~\cite{Gen2}.


\begin{acknowledgments}
The IceCube Collaboration designed, constructed and now operates the IceCube Neutrino Observatory. 
Data processing and calibration, Monte Carlo simulations of the detector and of theoretical models, 
and data analyses were performed by a large number of collaboration members, 
who also discussed and approved the scientific results presented here. 
The main author of this manuscript was Shigeru Yoshida. 
It was reviewed by the entire collaboration before publication, and 
all authors approved the final version of the manuscript.

We acknowledge the support from the following agencies:
USA -- U.S. National Science Foundation-Office of Polar Programs,
U.S. National Science Foundation-Physics Division,
Wisconsin Alumni Research Foundation,
Center for High Throughput Computing (CHTC) at the University of Wisconsin-Madison,
Open Science Grid (OSG),
Extreme Science and Engineering Discovery Environment (XSEDE),
U.S. Department of Energy-National Energy Research Scientific Computing Center,
Particle astrophysics research computing center at the University of Maryland,
Institute for Cyber-Enabled Research at Michigan State University,
and Astroparticle physics computational facility at Marquette University;
Belgium -- Funds for Scientific Research (FRS-FNRS and FWO),
FWO Odysseus and Big Science programmes,
and Belgian Federal Science Policy Office (Belspo);
Germany -- Bundesministerium f\"ur Bildung und Forschung (BMBF),
Deutsche Forschungsgemeinschaft (DFG),
Helmholtz Alliance for Astroparticle Physics (HAP),
Initiative and Networking Fund of the Helmholtz Association,
Deutsches Elektronen Synchrotron (DESY),
and High Performance Computing cluster of the RWTH Aachen;
Sweden -- Swedish Research Council,
Swedish Polar Research Secretariat,
Swedish National Infrastructure for Computing (SNIC),
and Knut and Alice Wallenberg Foundation;
Australia -- Australian Research Council;
Canada -- Natural Sciences and Engineering Research Council of Canada,
Calcul Qu\'ebec, Compute Ontario, Canada Foundation for Innovation, WestGrid, and Compute Canada;
Denmark -- Villum Fonden, Danish National Research Foundation (DNRF);
New Zealand -- Marsden Fund;
Japan -- Japan Society for Promotion of Science (JSPS)
and Institute for Global Prominent Research (IGPR) of Chiba University;
Korea -- National Research Foundation of Korea (NRF);
Switzerland -- Swiss National Science Foundation (SNSF).

\end{acknowledgments}



\begin{thebibliography}{99}
\bibitem{augerScience2017}
   A.~Aab {\it et al.} (Pierre Auger Collaboration), Science {\bf 357}, 1266 (2017).
\bibitem{EHE2011}
  R.~Abbasi {\it et al.} (IceCube Collaboration), Phys.~Rev.~D {\bf 83}, 092003 (2011).
\bibitem{EHE2012}
  M.~G.~Aartsen {\it et al.} (IceCube Collaboration), Phys.~Rev.~D {\bf 88}, 112008 (2013).
\bibitem{augerNu2015}
  A.~Aab {\it et al.} (Pierre Auger Collaboration), Phys. Rev. D {\bf 91}, 092008 (2015).
\bibitem{EHE2016}
  M.~G.~Aartsen {\it et al.} (IceCube Collaboration), Phys.~Rev.~Lett. {\bf 117}, 241101 (2016);
  M.~G.~Aartsen {\it et al.} (IceCube Collaboration), {\it ibid.} {\bf 119} 259902 (2017)
\bibitem{berezinsky69}
V.~S.~Berezinsky and G.~T.~Zatsepin, Phys.~Lett. B {\bf 28}, 423 (1969).
\bibitem{yoshidaIshihara2012}
S.~Yoshida and A.~Ishihara, Phys.~Rev. D {\bf 85}, 063002 (2012).
\bibitem{Kotera2010}
  K.~Kotera, D.~Allard, and A.~V.~Olinto, J.~Cosmol.~Astropart.~Phys. {\bf 2010}, 013 (2010).
\bibitem{Decerprit2011}
G.~Decerprit and D.~Allard, Astron.~Astrophys. {\bf 535}, A66 (2011).
\bibitem{augerICRC2017}
  Pierre Auger collaboration, Proceedings of Science {\bf 301} (ICRC 2017), 972 (2017);
  https://pos.sissa.it/301/972/
\bibitem{Anchordoqui}
  L.~A.~Anchordoqui et al., Phys.~Rev.~D {\bf 66}, 103002 (2002).
\bibitem{feldman98}
  G.~J.~Feldman and R.~D.~Cousins, Phys.~Rev.~D {\bf 57}, 3873  (1998).
\bibitem{icecubePeV2013}
  M.~G.~Aartsen {\it et al.} (IceCube Collaboration), Phys.~Rev.~Lett. {\bf 111}, 021103 (2013).
\bibitem{IceCubeDetector}
M.~G.~Aartsen {\it et al.} (IceCube Collaboration), J. Instrum. {\bf 12}, P03012 (2017).
\bibitem{corsika}
  D.~Heck {\it et al.}, {\it CORSIKA: a Monte Carlo code to simulate extensive air showers} (Forschungszentrum Karlsruhe, 1998), Report {\bf FZKA} 6019.
\bibitem{Ahn:2009wx}
E.-J. Ahn, R. Engel, T. K. Gaisser, P. Lipari, and T. Stanev, Phys. Rev.~D {\bf 80}, 094003 (2009).
\bibitem{anis}
A. Gazizov and M.~P. Kowalski, Comput.~Phys.~Commun. {\bf 172},  203  (2005).
\bibitem{Enberg2008}
R.~Enberg, M.~H.~Reno, and I.~Sarcevic, Phys. Rev. D {\bf 78}, 043005 (2008).
\bibitem{Bhattacharya2016}
A.~Bhattacharya {\it et al.}, J. High Energy Phys. {\bf 2016} 167 (2016);
R.~Gauld {\it et al.}, J. High Energy Phys. {\bf 2016} 130 (2016).
\bibitem{juliet}
  S.~Yoshida, R. Ishibashi, and H. Miyamoto, Phys.~Rev.~D {\bf 69}, 103004 (2004).
\bibitem{lineFit}
 R.~Abbasi {\it et al.} (IceCube Collaboration), Phys. Rev. D {\bf 82}, 072003 (2010).
\bibitem{EHElineFit}
 M.~G.~Aartsen {\it et al.} (IceCube Collaboration), Nucl. Instrum. Methods Phys. Res. A {\bf 736}, 143 (2014).
\bibitem{icecubeRealtime}
 M.~G.~Aartsen {\it et al.} (IceCube Collaboration), Astropart.~Phys. {\bf 92}, 30 (2017).
\bibitem{Ahlers2010}
M.~Ahlers {\it et al.}, Astropart. Phys. {\bf 34}, 106 (2010).
\bibitem{mrf}
G.~Hill and K.~Rawlins, Astropart.~Phys. {\bf 19}, 393 (2003).
\bibitem{IceCubeHESE2014}
M.~G.~Aartsen {\it et al.} (IceCube Collaboration), Phys.~Rev.~Lett. {\bf 113}, 101101 (2014).
\bibitem{reco}
  M.~G.~Aartsen {\it et al.} (IceCube Collaboration), J. Instrum. {\bf 9}, P03009 (2014).
\bibitem{spe12}
J. Ahrens et al. (AMANDA Collaboration), Nucl. Instrum. Methods Phys. Res. A {\bf 524}, 169 (2004).
\bibitem{diffuseNuMu6yr}
M.~G.~Aartsen {\it et al.} (IceCube Collaboration), Astrophys. J. {\bf 833}, no. 1, 3 (2016).
\bibitem{murase2014}
  K.~Murase, Y.~Inoue, and C.~D.~Dermer, Phys. Rev. D {\bf 90}, 023007 (2014).
\bibitem{anita}
P.~W.~Gorham {\it et al.} (ANITA Collaboration), Phys. Rev. D {\bf 98}, 022001 (2018).
\bibitem{EBL2017}
K.~Mattila {\it et al.}, Monthly Notices of the Royal Astronomical Society {\bf 470}, 2152 (2017).
\bibitem{Heinze2016}
J.~Heinze et al., Astrophys. J. {\bf 825}, no. 2, 122 (2016).
\bibitem{protonDip}
  V.~Berezinsky, A.~Gazizov, and S.~Grigorieva, Phys.~Rev.~D {\bf 74}, 043005 (2006).
\bibitem{Aloisio2015}
  R.~Aloisio {\it et al.}, J.~Cosmol.~Astropart.~Phys. {\bf 10}, 006 (2015).
\bibitem{SFR}
  A.~M.~Hopkins and J.~F.~Beacom, Astrophys.~J. {\bf 651}, 142 (2006);
  H.~Y\"{u}ksel {\it et al.}, Astrophys. J. Lett. {\bf 683}, L5 (2008).
\bibitem{BrezinskyFermi}
  V.~Berezinsky {\it et al.}, Phys. Lett. B {\bf 695}, 13 (2011);
  V.~Berezinsky, A.~Gazizov, and O.~Kalashev, Astropart.~Phys. {\bf 84}, 52 (2016).
\bibitem{globus2017}
  N.~Globus {\it et al.}, Astrophys.~J. {\bf 839}, L22 (2017).
\bibitem{fermi}
  M.~Ackermann {\it et al.} (Fermi-LAT Collaboration), Astrophys. J. {\bf 799}, 86 (2015).
\bibitem{ARA}
  P.~Allison {\it et al.} (ARA Collaboration), Phys. Rev. D {\bf 93}, 082003 (2016).
\bibitem{ARIANNA}
 S.~W.~Barwick {\it et al.} (ARIANNA Collaboration), Astropart. Phys. {\bf 70}, 12 (2015).
\bibitem{Gen2}
  IceCube collaboration, Proceedings of Science {\bf 301} (ICRC 2017), 991 (2017);
  https://pos.sissa.it/301/991/
\end{thebibliography}
\end{document}